\documentclass[11pt]{article}

\usepackage[T1]{fontenc}
\usepackage[utf8]{inputenc}
\usepackage[margin=1in]{geometry}
\usepackage{lmodern}
\usepackage{microtype}
\usepackage{graphicx}
\usepackage{amsmath,amssymb}
\usepackage{booktabs}
\usepackage{subcaption}
\usepackage{caption}
\usepackage{enumitem}
\usepackage{siunitx}
\usepackage[section]{placeins}
\usepackage[round,authoryear]{natbib}
\usepackage[colorlinks=true,linkcolor=blue,citecolor=blue,urlcolor=blue]{hyperref}

\setlist[enumerate]{leftmargin=*,itemsep=0.25em,topsep=0.25em}
\setlist[itemize]{leftmargin=*,itemsep=0.25em,topsep=0.25em}

\title{STEP-Parts: Geometric Partitioning of Boundary Representations for Large-Scale CAD Processing}
\author{%
Shen Fan$^{1}$, Miko\l aj Kida$^{2}$, Przemyslaw Musialski$^{1}$\\[0.5em]
\small $^{1}$New Jersey Institute of Technology\\
\small $^{2}$Warsaw University of Technology}
\date{}

\begin{document}

\maketitle

\begin{abstract}
Many CAD learning pipelines discretize Boundary Representations (B-Reps) into triangle meshes, discarding analytic surface structure and topological adjacency and thereby weakening consistent instance-level analysis. We present \textsc{STEP-Parts}, a deterministic CAD-to-supervision toolchain that extracts geometric instance partitions directly from raw STEP B-Reps and transfers them to tessellated carriers through retained source-face correspondence, yielding instance labels and metadata for downstream learning and evaluation. The construction merges adjacent B-Rep faces only when they share the same analytic primitive type and satisfy a near-tangent continuity criterion. On ABC, same-primitive dihedral angles are strongly bimodal, yielding a threshold-insensitive low-angle regime for part extraction. Because the partition is defined on intrinsic B-Rep topology rather than on a particular triangulation, the resulting boundaries remain stable under changes in tessellation. Applied to the DeepCAD subset of ABC, the pipeline processes approximately 180{,}000 models in under six hours on a consumer CPU. We release code and precomputed labels, and show that \textsc{STEP-Parts} serves both as a tessellation-robust geometric reference and as a useful supervision source in two downstream probes: an implicit reconstruction--segmentation network and a dataset-level point-based backbone.
\end{abstract}


\section{Introduction}\label{sec:intro}

\begin{figure*}[!b]
  \centering
  \includegraphics[width=0.99\textwidth]{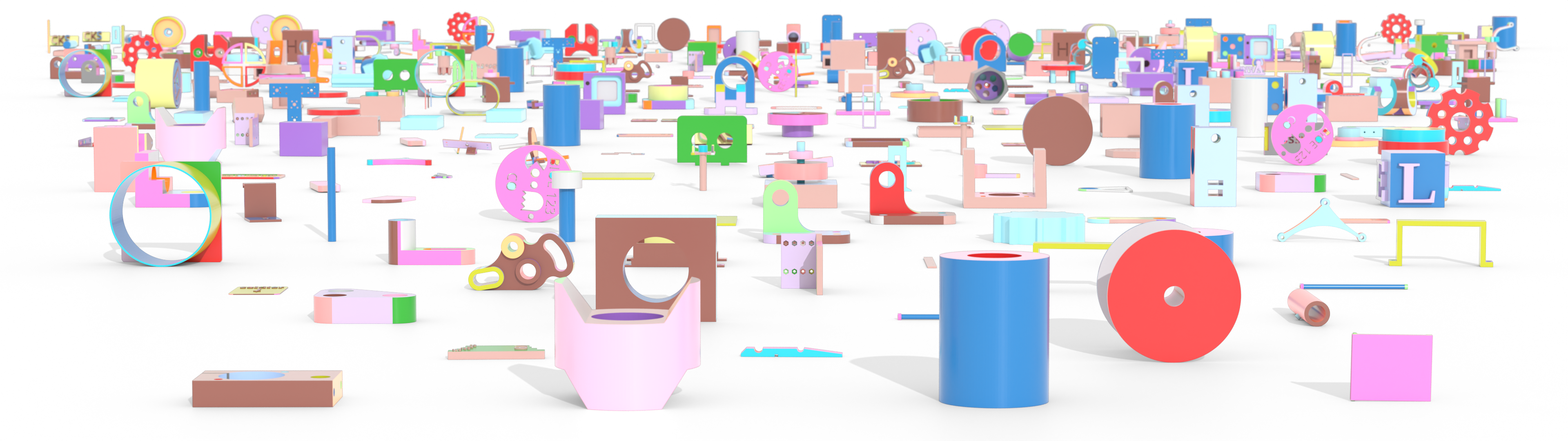}
  \caption{Rendered CAD models from the ABC/DeepCAD corpus, colored by STEP-Parts instances, illustrating our topology-aware B-Rep partitioning across thousands of geometrically diverse models.}
  \label{fig:teaser}
\end{figure*}

Geometric analysis of CAD models, including feature recognition, tolerance verification, and manufacturability assessment, relies on decomposing complex solids into coherent part-level structures \citep{BHANDARKAR20003,BABIC2008321}.
While parametric models are authored as Boundary Representations (B-Reps) containing analytic surfaces and explicit topological adjacency, standard processing pipelines routinely convert these models into discretized triangle meshes
\citep{Koch_2019_CVPR}.
This conversion discards the exact B-Rep structure, replacing intrinsic geometric boundaries with sampling artifacts and obscuring the instance-level definitions required for consistent analysis.

As a result, large-scale CAD corpora such as ABC \citep{Koch_2019_CVPR} lack reproducible, geometry-faithful part decompositions.
Mesh-based segmentation methods, while effective for generic 3D shapes, do not explicitly encode analytic surface structure or B-Rep topology.
At the same time, B-Rep-native learning methods such as BRepNet \citep{lambourne2021brepnet} and BRepFormer \citep{Dai2025BRepFormerTB} demonstrate the value of operating directly on topological CAD representations, but currently depend on unavailable or inconsistent ground-truth instance labels for training and evaluation. This mismatch between representation fidelity and available structural annotations limits both traditional geometric processing and emerging data-driven approaches on CAD data.

This motivates the following problem: how can raw STEP/B-Rep data be converted into reproducible, topology-aware instance supervision for CAD learning and evaluation? We address this gap with \textsc{STEP-Parts}, a deterministic CAD-to-supervision toolchain that derives topology-aware geometric instance labels directly from raw STEP B-Reps. The method groups adjacent B-Rep faces only when they share the same analytic primitive type and satisfy a near-tangent continuity criterion, and then transfers the resulting face-level partition to a tessellated carrier by retaining, for each triangle, the index of its source B-Rep face. Because the construction is defined on B-Rep topology and analytic surface classes rather than on a particular triangulation, it preserves intrinsic geometric discontinuities without relying on semantic taxonomies, design intent, or manual annotation. The resulting representation comprises triangle geometry, projected STEP-Part labels, source B-Rep face indices, primitive labels, and auxiliary metadata, and can be used directly in downstream learning and evaluation pipelines or rerun on other STEP corpora. The representation is also extensible: when required by downstream tasks, the same pipeline can be extended to attach additional analytic B-Rep attributes to the tessellated carrier.

This work makes the following contributions:
\begin{enumerate}
    \item \textbf{A deterministic, topology-aware CAD-to-supervision toolchain} that converts raw STEP/B-Rep models into geometric instance labels through primitive-aware, near-tangent merging on the B-Rep and transfer to tessellated carriers through retained source-face correspondence.

    \item \textbf{Empirical validation of a stable operating regime and a tessellation-robust geometric reference.} On ABC, same-primitive dihedral angles exhibit a strongly bimodal distribution with a sparse low-angle tail, yielding a stable regime for threshold-based merging. The selected operating point is calibrated for ABC, and the same procedure can be repeated on other STEP corpora. Because the partition is defined on the B-Rep, it exhibits substantially higher self-consistency across tessellation changes than a representative mesh-based alternative.

    \item \textbf{To our knowledge, the first large-scale topology-aware geometric instance-segmentation resource for STEP CAD,} instantiated on the DeepCAD subset of ABC with approximately 180{,}000 processed models and released together with open-source extraction code and derived labels.

    \item \textbf{Empirical evidence that STEP-Parts provides useful supervision for geometric learning,} demonstrated through two downstream probes under matched training conditions: a shape-level joint SDF--segmentation model ~\citep{Fan2025JointNS} and a dataset-level PTv3 point backbone \citep{Wu2023PointTV}.
\end{enumerate}

\section{Related Work}\label{sec:related}

\subsection{CAD Feature Recognition and Segmentation}

Geometric analysis of CAD models extends beyond simple shape classification to rigorous tasks such as tolerance verification, manufacturability assessment, and feature recognition~\citep{BHANDARKAR20003,BABIC2008321}. Traditional systems rely on explicit B-Rep graph traversal to identify semantic features (e.g., holes, slots) or functional regions. While recent learning-based approaches predict semantic labels on B-Rep faces~\citep{lambourne2021brepnet,Dai2025BRepFormerTB}, they typically depend on fixed machining taxonomies and do not address the fundamental problem of geometric instance segmentation. Our work targets this lower-level structural decomposition: we partition B-Rep solids into coherent geometric instances based purely on analytic primitive identity and topological continuity, providing a foundational representation independent of specific manufacturing intent.

\subsection{Deep Learning on B-Rep CAD Models}

Neural networks operating directly on B-Rep topology have emerged as a powerful paradigm for CAD analysis. Architectures such as BRepNet~\citep{lambourne2021brepnet} and BRepGAT~\citep{Lee2023BRepGATGN} utilize message passing on face-edge graphs for classification and segmentation, while Hierarchical CADNet~\citep{COLLIGAN2022103226} and BRepFormer~\citep{Dai2025BRepFormerTB} integrate geometric and topological features for complex recognition tasks. Hybrid methods like SpelsNet~\citep{Cherenkova2024SpelsNetSP} further combine B-Reps with point clouds. All these methods require supervision in the form of face-level annotations, which are scarce for large public datasets. Our work provides the necessary infrastructure to scale these approaches: by automatically deriving deterministic instance labels from raw STEP files, we enable B-Rep learning on large-scale unannotated corpora like ABC.

\subsection{Mesh-Based Part Segmentation}

A vast literature addresses part segmentation on discretized representations such as meshes and point clouds~\citep{Qi2017PointNetDH,Lian2020DeepMM}. Supervised methods typically learn semantic labels from fixed taxonomies (e.g., ShapeNetPart~\citep{Chang2015ShapeNetAI}), while unsupervised approaches like PartField~\citep{Liu2025PARTFIELDL3} cluster continuous 3D feature fields. Recent foundation models adapt Segment Anything~\citep{kirillov2023segany,ravi2024sam2} to 3D via multi-view rendering~\citep{Tang2024SegmentAM,Yang2024SAMPart3DSA}, while zero-shot methods like SATR~\citep{abdelreheem2023SATR} lift 2D predictions. However, because these methods operate on discretized geometry, they do not explicitly enforce topological validity at analytic boundaries. Our B-Rep-derived segmentation serves as a deterministic, topology-preserving reference, enabling rigorous evaluation of how well mesh-based methods recover intrinsic CAD structure.

\subsection{Part-Aware Generative Models}

Part-level decomposition is essential for controllable 3D synthesis and assembly. Generative frameworks such as PartGen~\citep{Chen2024PartGenP3}, PartCrafter~\citep{lin2025partcrafter}, and Assembler~\citep{Zhao2025AssemblerS3} rely on part-structured assets like PartNet~\citep{Mo2018PartNetAL} to produce editable models. Similarly, joint reconstruction-segmentation networks~\citep{Fan2025JointNS} demonstrate that part supervision improves implicit surface reconstruction. Training these models requires consistent, geometrically valid part decompositions. By extracting parts directly from the B-Rep, we provide a training signal that respects the analytic nature of CAD data, offering a more rigorous basis for generative modeling than heuristic mesh clustering.

\subsection{CAD Datasets and Benchmarks}

The ABC dataset~\citep{Koch_2019_CVPR} provides approximately one million STEP models but lacks instance-level part annotations, limiting its utility for part-aware analysis. DeepCAD~\citep{Wu_2021_ICCV} offers a curated subset with modeling sequences but no explicit B-Rep part labels. Other resources, such as the Fusion 360 Gallery~\citep{willis2020fusion,willis2021joinable}, specialized machining datasets~\citep{ZHANG2024102318}, or mesh-based resources like PartObjaverse-Tiny~\citep{Yang2024SAMPart3DSA} on Objaverse~\citep{objaverse}, are often restricted in scope or availability. While tools like Better STEP~\citep{Izadyar2025BetterSA} facilitate B-Rep processing, they do not supply large-scale instance labels. Our work fills this gap by converting raw ABC STEP files into per-triangle instance labels that reflect geometric and topological structure without relying on machining taxonomies, operation logs, or manual annotation, establishing a reproducible benchmark for the community.

\section{STEP-Parts Extraction}
\label{sec:method}

We define \textsc{STEP-Parts} as a deterministic geometric partition of B-Rep faces and realize this definition as a CAD-to-supervision toolchain. Starting from a raw STEP model, we (i) tessellate each B-Rep face under fixed geometric tolerances, (ii) assign each face an analytic primitive type, (iii) build the exact face-adjacency graph from STEP topology, (iv) merge adjacent faces only when they are primitive-compatible and near-tangent, and (v) transfer the resulting face-level partition to a tessellated carrier by retaining, for each triangle, the index of its source B-Rep face. The result is a carrier-level representation comprising triangle geometry, projected STEP-Part labels, source B-Rep face indices, primitive labels, and auxiliary metadata for visualization, evaluation, and downstream learning.

\begin{figure*}[t]
  \centering
  \includegraphics[width=0.99\textwidth]{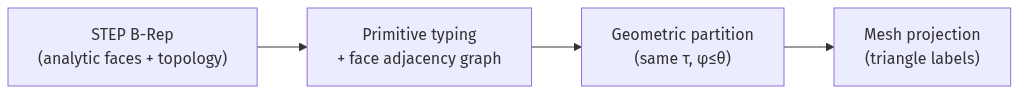}
  \caption{Overview of the STEP-Parts extraction pipeline. Starting from a STEP B-Rep (analytic surface faces with explicit topology), we assign each face an analytic primitive type and build the face adjacency graph, then form STEP-Parts as connected components under the geometric predicate same primitive type $t(f)$ and edge-anchored dihedral $\phi \le \theta$. The resulting face-level partition is then transferred to a tessellated carrier by retaining, for each triangle, the index of its source B-Rep face and assigning the corresponding STEP-Part identifier, yielding per-triangle instance labels.}
  \label{fig:pipeline_widescreen}
\end{figure*}

\subsection{B-Rep Structure and Geometric Domain}

\noindent
A B-Rep solid is represented as a tuple
\[
\mathcal{B} = (F, E_B, \Sigma),
\]
where $F=\{f_1,\dots,f_n\}$ is a finite set of faces, $E_B$ is a set of topological edges, and each face $f \in F$ is associated with an analytic surface $\sigma(f) \in \Sigma$ together with a trimming that defines a bounded surface patch. 
The mapping $\sigma : F \rightarrow \Sigma$ assigns to each face its underlying analytic surface, defined by a primitive type and associated continuous parameters (e.g., axis, radius). 
Two faces $f_a, f_b \in F$ are adjacent if and only if they share a topological edge $e \in E_B$.

Importantly, B-Rep faces are not discrete facets but trimmed regions of continuous analytic surfaces. Trimming operations may subdivide an analytic surface into multiple faces without introducing a geometric discontinuity; STEP-Parts therefore treat trimming boundaries as topological artifacts rather than part boundaries. Multiple faces can thus correspond to the same analytic surface $\sigma$ while remaining topologically distinct. This separation between analytic surface identity and topological subdivision is fundamental to the construction of STEP-Parts.

\subsection{Geometric Partitioning of B-Rep Faces}

We formulate STEP-Part extraction as defining an equivalence relation over the face set $F$.
Two faces are considered part of the same geometric instance if they cannot be distinguished by local intrinsic geometric properties. 
At the B-Rep level, analytic primitive type provides the strongest intrinsic distinction between faces, as ignoring it cannot capture curvature-class changes introduced by modeling operations. 
Within a single primitive class, local tangent continuity is the remaining intrinsic criterion.

Formally, we define two adjacent faces $f_a, f_b \in F$ to be equivalent if and only if
\[
t(f_a) = t(f_b) \ \text{and} \ \phi(f_a,f_b) \le \theta,
\]
where $t(f)$ denotes the analytic primitive type of face $f$ and $\phi(f_a,f_b)$ is the dihedral angle
evaluated at their shared topological edge. 
The threshold $\theta$ is treated as an operating setting rather than a tuned per-shape hyperparameter. In the experiments below, we use $\theta = 8.0^\circ$ as the ABC operating point, with empirical justification deferred to Sec.~\ref{subsec:dihedral_hist}. Primitive homogeneity ensures analytic compatibility, while near-tangent continuity preserves smooth surface transitions and isolates intrinsic geometric discontinuities without invoking semantic intent or design history. The same calibration procedure can be repeated for other STEP corpora when their same-primitive dihedral statistics differ materially from those observed on ABC.

Tangent continuity is evaluated \emph{at the shared interface}. For two adjacent faces sharing a topological edge $e\in E_B$, we compute oriented surface normals $n_a$ and $n_b$ at a common parameter value along $e$ and define
\[
\phi = \arccos\!\big(\operatorname{clamp}(n_a \cdot n_b,\,-1,\,1)\big).
\]
Evaluating normals along the shared edge distinguishes smooth ($C^1$) joins from sharp features intrinsically and avoids ambiguity on trimmed or curved surfaces. Face orientation is handled explicitly to ensure consistent normal alignment.

This local equivalence relation induces a global decomposition of the solid. The B-Rep face adjacency graph $G=(F,E)$ encodes exact topology, and STEP-Parts are obtained as the connected components of $G$ under the equivalence predicate above, computed via breadth-first flood-fill. The algorithmic realization is therefore a direct consequence of the geometric definition.

\subsection{Tessellation, Carrier Representation, and Stabilization}

The STEP-Part partition is defined entirely on the B-Rep structure $\mathcal{B}$. Tessellation is introduced only to provide a discrete carrier for visualization, evaluation, and downstream learning. Each B-Rep face is tessellated independently under fixed geometric tolerances, and each resulting triangle retains the index of its source B-Rep face. If $\ell : F \to \{1,\ldots,k\}$ denotes the face-level STEP-Part assignment and $f_j$ is the source face of triangle $j$, then the carrier-level label is
\[
y_j = \ell(f_j).
\]
We additionally replicate the analytic primitive type of each source face to its triangles. The resulting carrier-level representation therefore consists of triangle geometry, projected STEP-Part labels, source B-Rep face indices, primitive labels, and auxiliary metadata.

Because the partition is defined prior to discretization, its boundaries are intended to reflect intrinsic B-Rep geometry rather than meshing artifacts. In practice, trimming and tessellation may still introduce very small carrier fragments. We therefore apply a minimum-triangle post-processing rule ($\tau_{\min}=20$ in our implementation) to suppress spurious micro-parts and compact the carrier-level labels. This step serves as representation hygiene: it stabilizes the tessellated carrier without changing the underlying B-Rep predicate.

The carrier representation is also intentionally extensible. Beyond the labels used in this paper, the same source-face correspondence can be used to attach additional analytic B-Rep attributes when required by downstream tasks, such as sampled surface points and normals, curvature-derived quantities, UV coordinates, principal directions, or face-adjacency information.

\subsection{Interpretation and Scope}

\textsc{STEP-Parts} inherit the modeling structure of the underlying B-Rep. Regions created as a single analytic primitive---for example, as part of an extrusion or sweep---remain grouped even when they appear visually composite, whereas distinct primitives remain separate regardless of scale. The resulting decomposition should therefore be interpreted as a topology-faithful geometric prior rather than a semantic or functional part taxonomy. It is suitable as a deterministic reference for CAD evaluation and as supervision for geometry-aware learning, but it does not guarantee that every mechanically meaningful component is isolated.

\section{Validation}
\label{sec:valid}

\begin{figure*}[t]
    \centering
    \begin{minipage}[t]{0.46\textwidth}
        \centering
        \includegraphics[width=\linewidth,trim=40 25 40 30,clip]{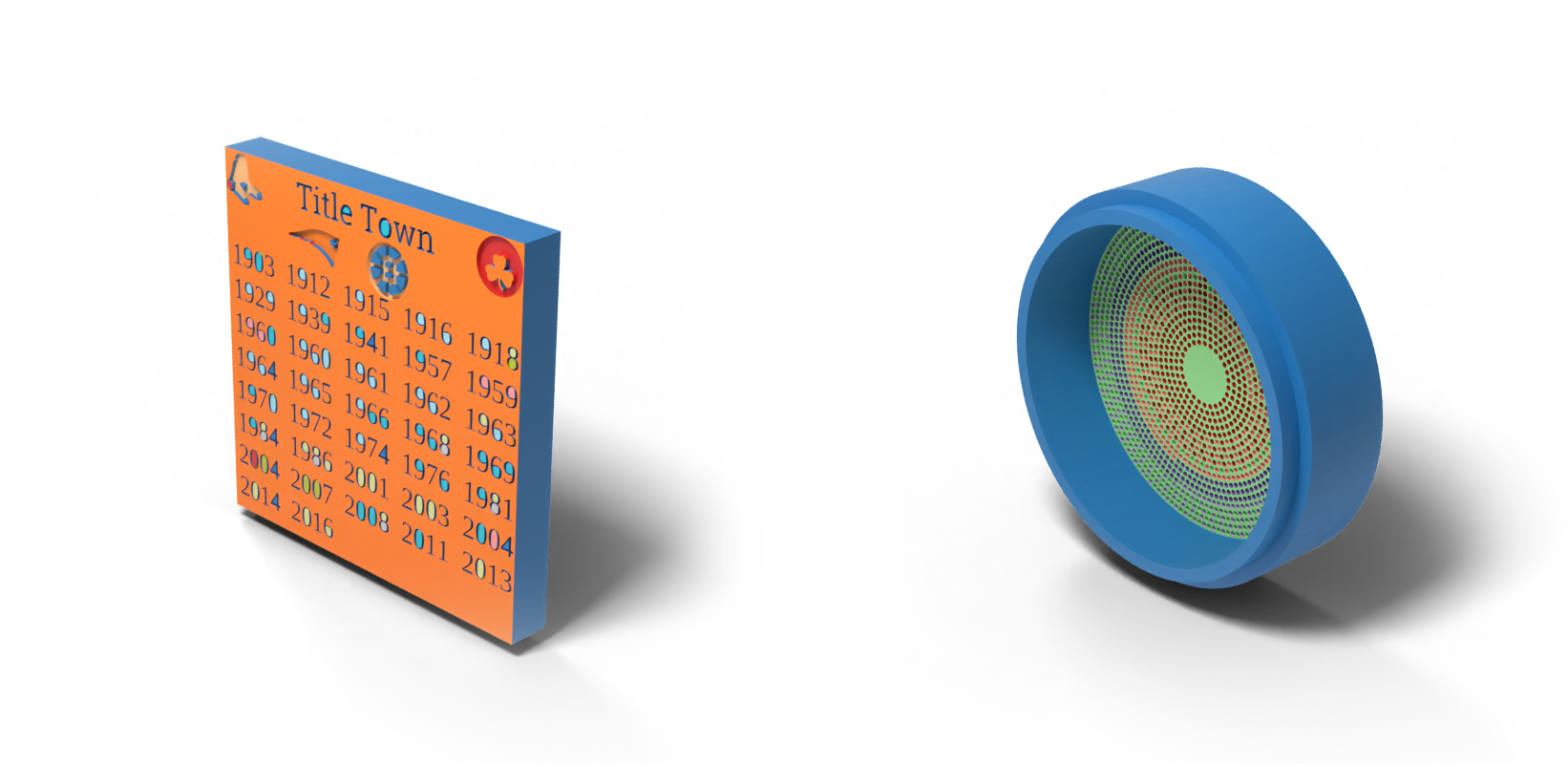}
        \captionof{figure}{Examples of CAD models from the DeepCAD subset with a large number of STEP-based parts. Our deterministic B-Rep-driven pipeline can extract such partitions without GPU acceleration.}
        \label{fig:step_many_parts}
    \end{minipage}\hfill
    \begin{minipage}[t]{0.52\textwidth}
        \centering
        \includegraphics[width=\linewidth]{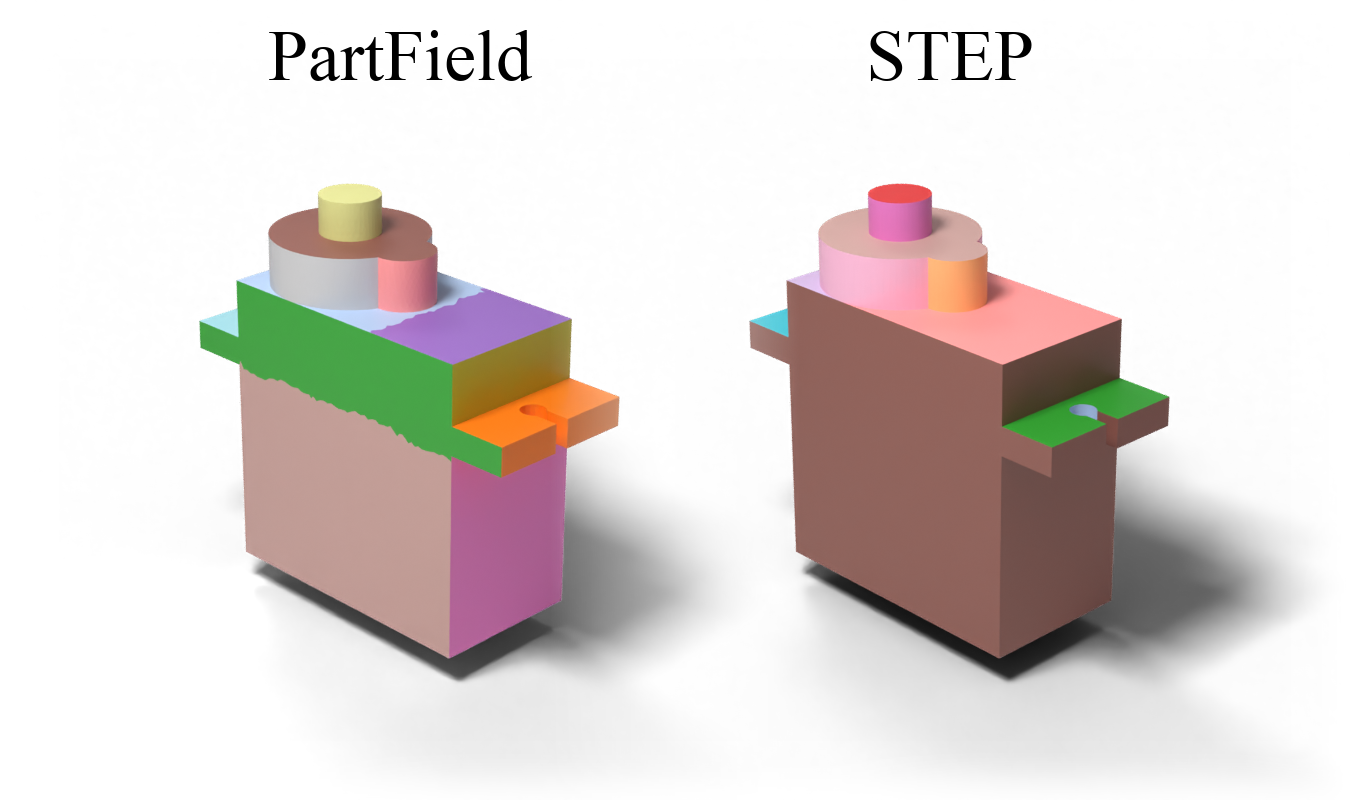}
        \captionof{figure}{Qualitative comparison of segmentation boundaries on ABC models. \textbf{Left:} PartField (mesh-based segmentation). \textbf{Right:} STEP-Parts (B-Rep-driven).}
        \label{fig:vis_pf_vs_step}
    \end{minipage}
\end{figure*}

This section validates the geometric construction defined in Sec.~\ref{sec:method}. The experiments are designed to assess three properties implied by the definition of STEP-Parts: (i) whether the tangent-continuity predicate is empirically well posed on real CAD data, (ii) whether the induced partition is invariant under discretization, and (iii) whether the construction is practical at scale. We further evaluate these properties through comparison with PartField~\citep{Liu2025PARTFIELDL3}, a learned part segmentation method, and through two complementary downstream learning probes: an implicit joint reconstruction--segmentation network~\citep{Fan2025JointNS} and Point Transformer V3 (PTv3)~\citep{Wu2023PointTV}, a dataset-level point-based backbone.

\subsection{Implementation and Evaluation Protocol}
\label{subsec:protocol}
All experiments use the same STEP-based preprocessing pipeline described in Sec.~\ref{sec:method}. For each CAD model, STEP-Parts are extracted on the B-Rep and then transferred to a tessellated carrier by retaining, for each triangle, the index of its source B-Rep face. The resulting carrier-level data include mesh geometry, projected STEP-Part labels, source B-Rep face indices, primitive labels, and auxiliary metadata. Unless stated otherwise, all validation experiments operate on this common carrier representation; any probe-specific alignment or label-transfer steps are described in the corresponding subsection.

\textbf{Datasets.}
Validation experiments are conducted on subsets of the ABC dataset following the DeepCAD selection. For comparisons to mesh-based methods, we restrict attention to models for which both a valid STEP file and a corresponding triangle mesh are available and successfully processed by PartField. This yields a validation set of 240 models used consistently across comparative experiments.

\textbf{Sampling and label transfer.}
For STEP-mesh comparisons, we uniformly sample $10^5$ points on the tessellated STEP surface of each model and assign each point a STEP-Part label according to its originating face. Mesh-based labels are transferred to the same points using a nearest-neighbor query on mesh vertices, providing a stable correspondence without requiring face-face intersection tests.

\textbf{Label alignment.}
Because STEP-Part identifiers and mesh-based labels are unordered, we align labels using an optimal permutation prior to evaluation. For each shape, we construct a confusion matrix between STEP labels and mesh-based labels over sampled points and compute a maximum-weight matching via the Hungarian algorithm. All reported accuracy and IoU metrics are computed after this alignment.

\textbf{Metrics.}
We report overall point-wise accuracy, mean intersection-over-union (mIoU) over STEP-Part instances, and boundary accuracy evaluated on points sampled in a narrow neighborhood of STEP-Part boundaries.

\begin{figure}[t]
  \centering
  \includegraphics[width=\linewidth]{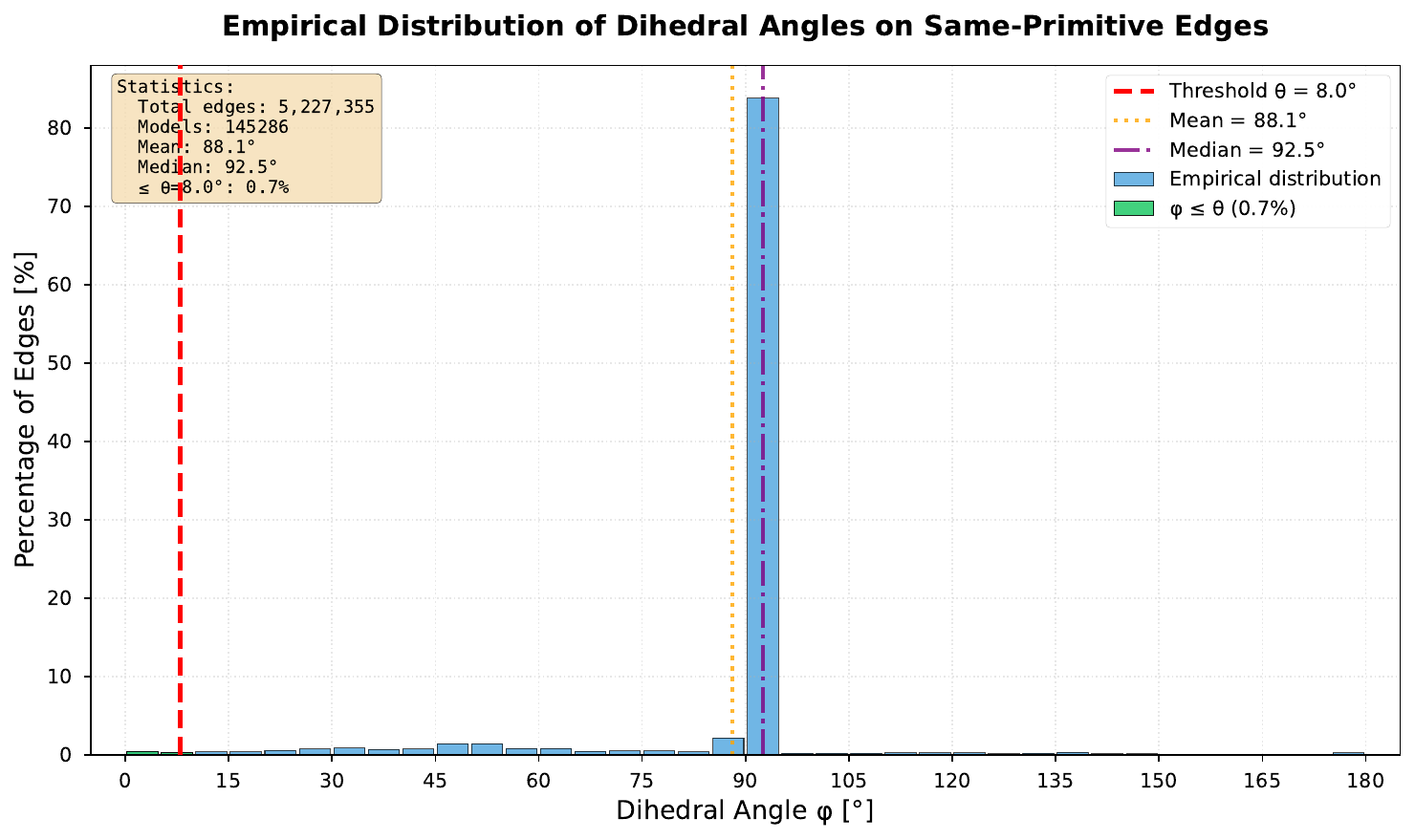}
  \caption{Global distribution of dihedral angles $\phi$ over all same-primitive face adjacencies in our ABC/DeepCAD subset. The mass concentrates near $\phi\approx 90^\circ$ (sharp features), with a sparse near-zero tail (tangent-continuous joins / small numerical effects). Some models contain no same-primitive adjacencies.}
  \label{fig:dihedral_hist_global}
\end{figure}

\subsection{Empirical Structure of Same-Primitive Dihedral Angles}
\label{subsec:dihedral_hist}

The STEP-Parts definition relies on a local tangent-continuity predicate, evaluated via the dihedral angle $\phi$ between adjacent faces that share the same analytic primitive type. To assess whether this predicate is well posed on real CAD data, we compute $\phi$ over all same-primitive face adjacencies in our ABC/DeepCAD subset.

Figure~\ref{fig:dihedral_hist_global} shows that the distribution is dominated by a sharp concentration near right angles, together with a sparse near-zero tail. In particular, only a small fraction of same-primitive adjacencies fall below our operating threshold (in our corpus, $\phi \le 8^\circ$ accounts for $0.7\%$ of edges), whereas the bulk of mass lies near $\phi \approx 90^\circ$. This empirical structure supports using a single global threshold within the low-angle region: $\theta$ primarily selects the rare tangent-continuous connections and leaves the dominant sharp-angle structure unaffected. Operationally, the threshold is needed to decide which same-primitive face adjacencies should be merged into a single STEP-Part and which should remain part boundaries. On ABC, the bimodal dihedral structure makes this decision insensitive within the tested low-angle regime; for other STEP corpora, the same calibration procedure can be rerun if their same-primitive dihedral statistics differ materially.

\paragraph{Local Stability of the Partition Around $\theta = 8^\circ$}
While the global histogram characterizes the available dihedral structure, we also verify that the induced partition is stable under small perturbations of $\theta$ around the operating point. We rerun extraction with $\theta \in \{4^\circ, 6^\circ, 8^\circ, 10^\circ, 12^\circ\}$ on a fixed subset of models.

\begin{figure}[t]
    \centering
    \includegraphics[width=0.9\linewidth]{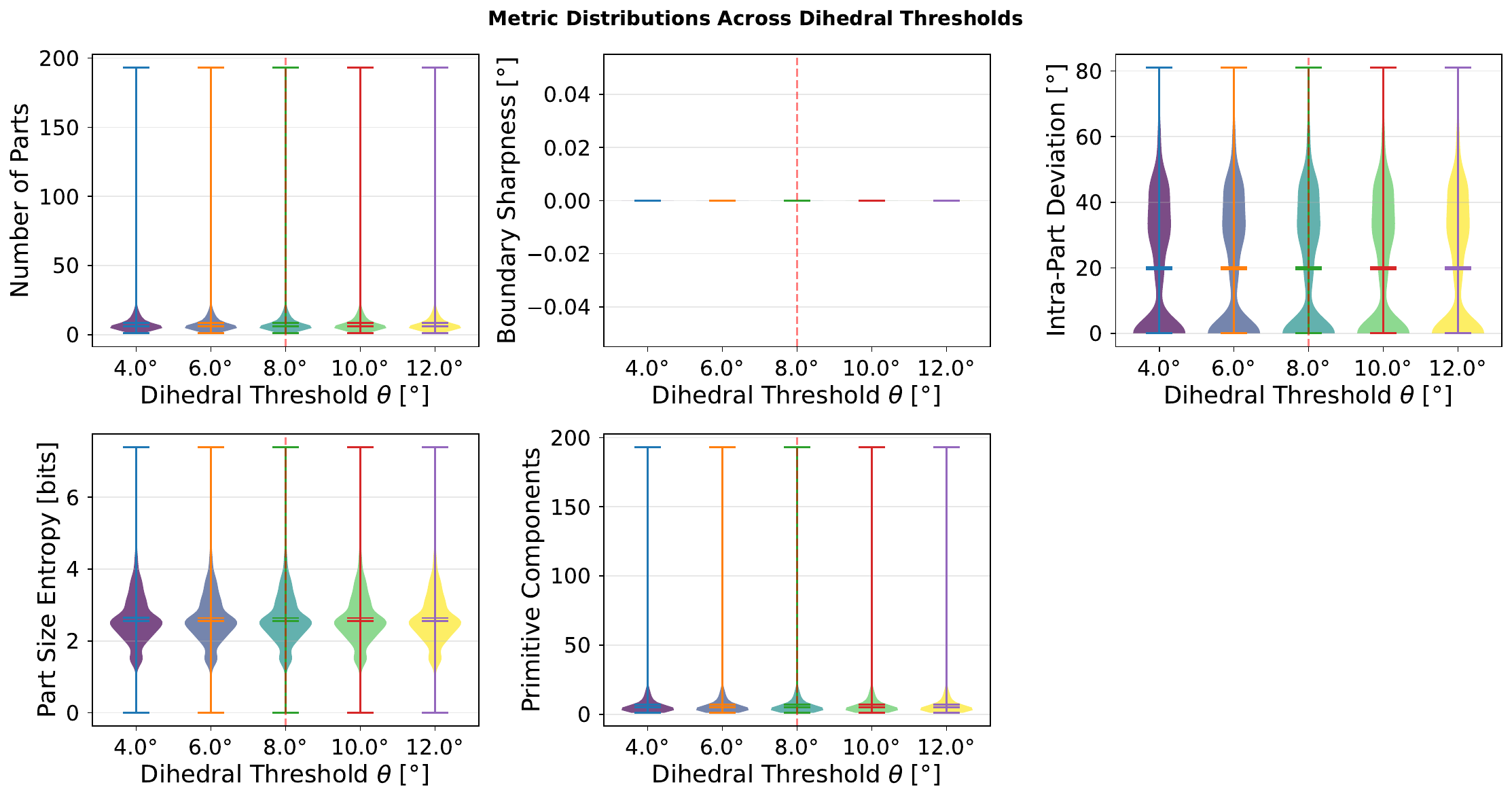}
    \caption{Local stability of STEP-Parts under threshold perturbations. Partition statistics and auxiliary quality metrics remain effectively unchanged for $\theta \in \{4^\circ,6^\circ,8^\circ,10^\circ,12^\circ\}$, indicating a stable low-angle operating regime rather than sensitivity to a finely tuned threshold.}
    \label{fig:supp_theta_sweep_metrics}
\end{figure}
Figure~\ref{fig:supp_theta_sweep_metrics} shows that the distributions of part count, boundary sharpness, intra-part deviation, part-size entropy, and primitive component counts remain nearly unchanged across the tested thresholds.

\begin{figure}[t]
    \centering
    \includegraphics[width=0.9\linewidth]{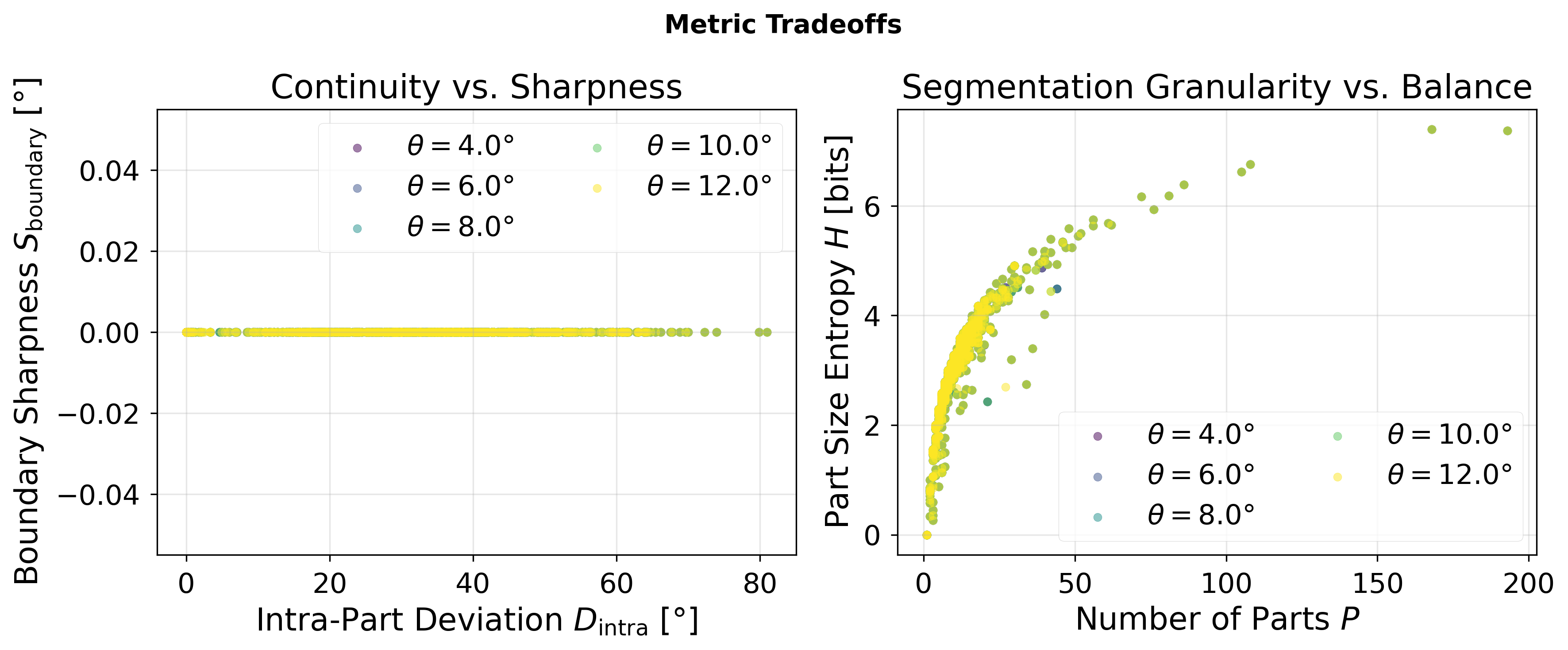}
    \caption{Metric tradeoffs under dihedral-threshold perturbations. Each point is one shape; colors indicate the threshold $\theta \in \{4^\circ,6^\circ,8^\circ,10^\circ,12^\circ\}$. Left: intra-part deviation $D_{\mathrm{intra}}$ versus boundary sharpness $S_{\mathrm{boundary}}$. Right: part-count $P$ versus part-size entropy $H$. The near-complete overlap of colors indicates that these aggregate metrics are effectively invariant to $\theta$ in the tested low-angle regime.}
    \label{fig:supp_tradeoffs}
\end{figure}

Figure~\ref{fig:supp_tradeoffs} provides the same conclusion from a complementary viewpoint: the strong overlap across colors in the pairwise metric plots indicates that the aggregate operating point of the partition remains essentially unchanged over the tested interval. Together, these results indicate that STEP-Parts does not rely on finely tuned threshold selection; we therefore use $\theta = 8^\circ$ as the ABC operating point within a stable low-angle regime.

\subsection{Tessellation Invariance}
\label{subsec:tessellation}

A central design goal of STEP-Parts is that the partition be defined on B-Rep topology and analytic surfaces rather than on a particular discretization. We therefore evaluate the stability of the extracted partitions under changes in tessellation resolution.

For each shape in the 240-model validation set, we generate multiple tessellations with varying chordal and angular tolerances and transfer the B-Rep--defined STEP-Parts labels to each tessellated carrier through retained source B-Rep face indices, applying the same carrier-level post-processing (including $\tau_{\min}$) before measuring self-consistency. Using a reference tessellation, we measure self-consistency across tessellations via label agreement and mean IoU after optimal alignment. The results are summarized in Table~\ref{tab:tessellation_self_consistency}. 

STEP-Parts exhibits very high self-consistency across both finer and coarser tessellations, with self-accuracy close to 0.99 and self-mIoU above 0.93 even under substantially coarser discretization. In contrast, mesh-based segmentations show lower self-consistency and greater variance, particularly near part boundaries. These results confirm that the STEP-Part boundaries reflect intrinsic B-Rep geometry rather than discretization artifacts. Refer to Figure~\ref{fig:tess_comp} for visual comparison. 

\noindent
\textit{Remark: } %
Across tessellation levels, the projected instance labels may differ in cardinality due to the discrete-carrier post-processing step that absorbs components with fewer than $\tau_{\min}=20$ triangles (Sec.~3.3), which can remove extremely thin regions under coarser tessellations.
For evaluation, we therefore align labels between $T_0$ and $T_j$ via maximum-weight matching (Hungarian algorithm); any unmatched instances are treated as mismatches (zero overlap) and thus lower self-mIoU, so the reported scores already include the penalty for missing parts.
For PartField, the number of predicted clusters is fixed by the clustering budget (e.g., capped at 20), which can yield stable cardinality by construction but does not imply tessellation-invariant boundaries.

\begin{figure}[t]
  \centering
  \includegraphics[width=\linewidth]{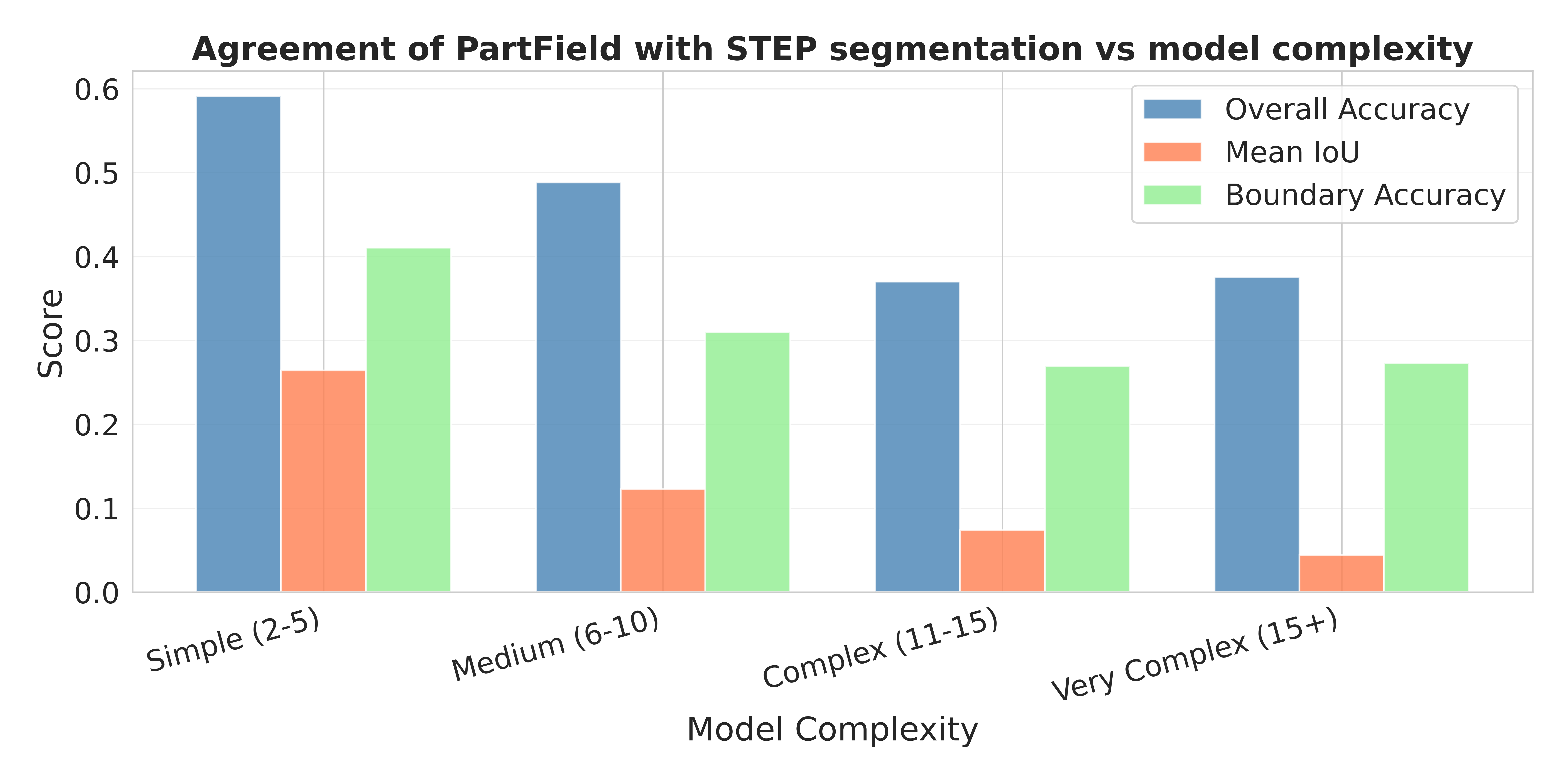}
  \caption{Agreement of PartField with our STEP-based segmentation as a function of model complexity (measured by the number of STEP parts). Bars show mean overall accuracy, mean IoU, and boundary accuracy of PartField with respect to the STEP reference, grouped by complexity bin. (Sec.~\ref{subsec:pf_comparison})}
  \label{fig:quality_by_complexity}
\end{figure}

\begin{table*}[b]
  \centering
  \setlength{\tabcolsep}{4pt}
  \small

  \begin{minipage}[b]{0.49\textwidth}
    \centering
    \caption{Global agreement between PartField segmentations and the STEP-based reference on the ABC dataset. We report mean and standard deviation for overall accuracy, mean IoU computed over STEP parts, and boundary accuracy.}
    \label{tab:pf_global}
    \begin{tabular*}{\linewidth}{@{\extracolsep{\fill}}lcc}
      \toprule
      Metric & Mean & Std \\
      \midrule
      Overall accuracy              & 0.43 & 0.17 \\
      Mean IoU (over STEP parts)    & 0.10 & 0.10 \\
      Boundary accuracy             & 0.30 & 0.12 \\
      \bottomrule
    \end{tabular*}
  \end{minipage}\hfill
  \begin{minipage}[b]{0.49\textwidth}
    \centering
    \caption{Tessellation self-consistency of STEP-Parts and PartField on 240 ABC shapes. For each method and tessellation pair $(T_0,T_j)$, we report self-accuracy and self-mIoU between the default tessellation $T_0$ and a finer ($T_1$) or coarser ($T_2$) tessellation.}
    \label{tab:tessellation_self_consistency}
    \begin{tabular*}{\linewidth}{@{\extracolsep{\fill}}lccccc}
      \toprule
      Method & Pair & $N$ & Self-Acc. & Self-mIoU \\
      \midrule
      PartField   & $T_0\!\leftrightarrow\!T_1$ & 240 & 0.94 $\pm$ 0.08 & 0.89 $\pm$ 0.13 \\
      PartField   & $T_0\!\leftrightarrow\!T_2$ & 240 & 0.94 $\pm$ 0.08 & 0.89 $\pm$ 0.12 \\
      STEP--Parts & $T_0\!\leftrightarrow\!T_1$ & 240 & 0.99 $\pm$ 0.04 & 0.99 $\pm$ 0.03 \\
      STEP--Parts & $T_0\!\leftrightarrow\!T_2$ & 240 & 0.99 $\pm$ 0.04 & 0.93 $\pm$ 0.12 \\
      \bottomrule
    \end{tabular*}
  \end{minipage}
\end{table*}

\subsection{Comparison to PartField}
\label{subsec:pf_comparison}

To evaluate the implications of defining parts on the B-Rep rather than on discretized geometry, we compare STEP-Parts to a representative segmentation method, PartField~\citep{Liu2025PARTFIELDL3}, on the 240-model validation set for which both methods successfully process the same shape instances. All comparisons use the point-sampling and optimal label alignment protocol described in Sec.~\ref{subsec:protocol}.

\begin{figure*}[!t]
  \centering
  \begin{minipage}[t]{0.49\textwidth}
    \centering
    \includegraphics[width=\linewidth]{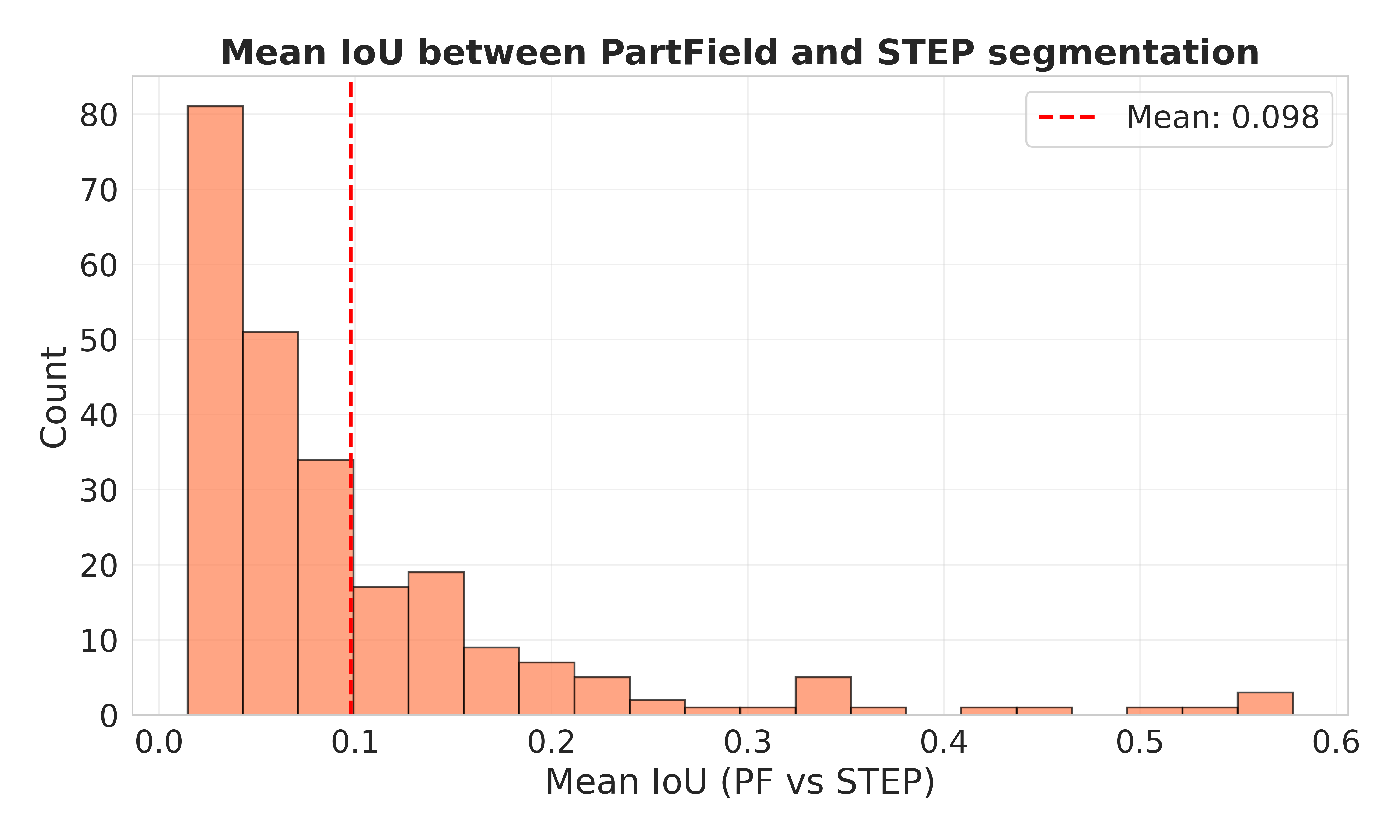}
    \caption{Distribution of per-shape mean IoU between PartField and the STEP-based segmentation after optimal label permutation (Sec.~\ref{sec:valid}). The dashed line marks the global mean ($\approx 0.10$), indicating that most shapes exhibit limited part-wise alignment.}
    \label{fig:iou_distribution}
  \end{minipage}\hfill%
  \begin{minipage}[t]{0.49\textwidth}
    \centering
    \includegraphics[width=\linewidth]{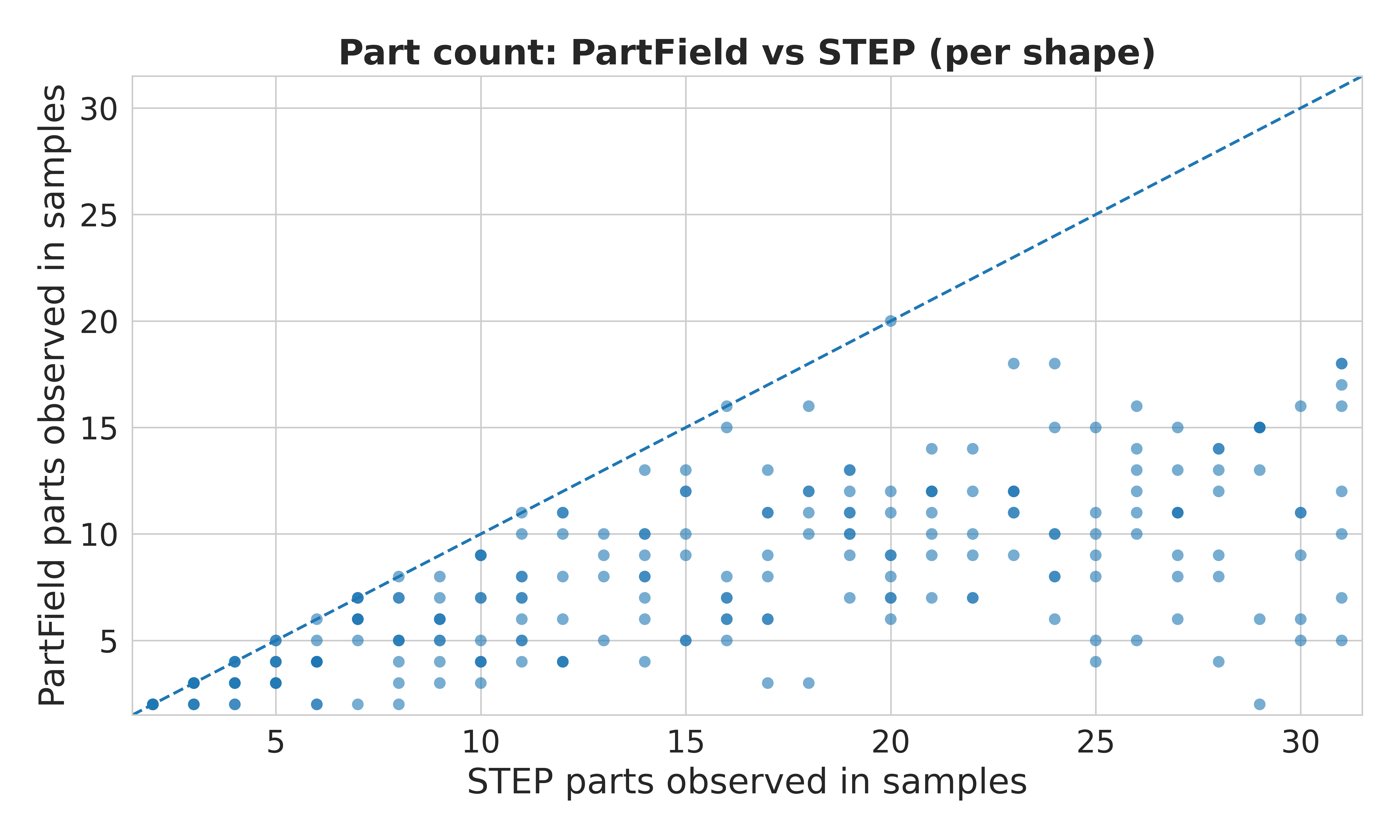}
    \caption{Per-shape comparison of the number of parts observed in samples for the STEP-based segmentation and the PartField segmentation. Each point corresponds to one STEP/PartField pair; the dashed line indicates perfect agreement.}
    \label{fig:part_count_comparison}
  \end{minipage}
\end{figure*}

\paragraph{Global agreement}
After optimal label alignment, PartField exhibits limited agreement with the STEP-based reference in terms of instance-level overlap and boundary localization (Table~\ref{tab:pf_global}). Figure~\ref{fig:iou_distribution} further shows that the per-shape mean IoU distribution is strongly concentrated at low values, with a long tail of better-matching cases, indicating that close part-wise alignment occurs only for a minority of shapes under this evaluation setting.

\paragraph{Dependence on model complexity}
Agreement degrades systematically with geometric complexity when complexity is measured by the number of STEP parts (Fig.~\ref{fig:quality_by_complexity}). While overall accuracy decreases with complexity, the strongest drop is observed in part-wise overlap (mean IoU), consistent with increasing granularity of the STEP-based partition on more complex shapes.

\paragraph{Part-count mismatch and qualitative behavior}
Part-count analysis (Fig.~\ref{fig:part_count_comparison}) shows that PartField typically predicts fewer parts than STEP-Parts on the same shapes, with saturation effects consistent with a fixed part-budget configuration. This under-partitioning contributes directly to low instance IoU values, since multiple STEP instances are merged into fewer mesh segments. Qualitative boundary comparisons (Fig.~\ref{fig:vis_pf_vs_step}) suggest that discrepancies are concentrated near analytic boundaries and thin regions, where mesh-based boundaries tend to be less aligned with the intrinsic B-Rep structure.

Overall, these results highlight the practical difference between a topology- and analytic-structure-driven partition defined on the B-Rep and a segmentation inferred on a discretized carrier under fixed capacity constraints, and motivate STEP-Parts as a stable geometric reference for part-level CAD analysis.

These discrepancies reflect both geometric and practical factors: PartField operates on a discretized carrier and deviates most strongly near analytic boundaries and thin regions, while comparison on ABC is further constrained by missing original meshes, GPU-heavy processing, and the visible-label ceiling of the released exporter.

The preceding experiments assess STEP-Parts as a geometry-native reference. We now turn to a complementary question: whether the same labels also provide useful supervision for downstream geometric learning.

\subsection{Implicit Reconstruction--Segmentation Probe}
\label{sec:implicit_probe}

We first assess STEP-Parts as a source of supervision using the joint SDF--segmentation network~\citep{Fan2025JointNS}, which augments an implicit SDF reconstruction trunk with a segmentation head. In contrast to dataset-level point-cloud backbones, this probe is used here as a shape-level implicit reconstruction-and-segmentation model; accordingly, we restrict the evaluation to a subset of 25 ABC meshes due to the substantially higher per-shape optimization cost.

Two instances of the network are trained under identical settings, differing only in the source of part supervision. Both models achieve comparable reconstruction quality, while STEP-Parts supervision yields higher segmentation accuracy, consistency, and part-wise agreement (Table~\ref{tab:joint_step_vs_pf}). These results indicate that STEP-Parts labels are locally coherent and compatible with implicit geometry-aware learning pipelines, while not degrading the reconstruction behavior of the underlying SDF model.

\begin{table*}[t]
  \centering
  \caption{Downstream joint SDF--segmentation evaluation on 25 ABC meshes. Both models share the same network and training settings and differ only in the source of segmentation supervision. Means and standard deviations are reported over shapes; arrows indicate the desired direction.}
  \label{tab:joint_step_vs_pf}
    \setlength{\tabcolsep}{4pt}
    \small
  \begin{tabular*}{\textwidth}{@{\extracolsep{\fill}}lccccccc}
    \toprule
    Supervision & $N$ &
    CDL1 ($\times 10^2$)$\downarrow$ & F1$_\mu\uparrow$ & NC$\uparrow$ &
    mIoU$\uparrow$ & Acc.$\uparrow$ & Cons.$\uparrow$ \\
    \midrule
    PartField &
    25 &
    0.53$\ \pm$0.59 &
    0.83$\ \pm$0.26 &
    0.96$\ \pm$0.06 &
    0.89$\ \pm$0.12 &
    0.95$\ \pm$0.08 &
    0.95$\ \pm$0.02 \\
    STEP-Parts &
    25 &
    \textbf{0.38}$\ \pm$0.38 &
    \textbf{0.91}$\ \pm$0.15 &
    \textbf{0.98}$\ \pm$0.02 &
    \textbf{0.91}$\ \pm$0.12 &
    \textbf{0.99}$\ \pm$0.04 &
    \textbf{0.98}$\ \pm$0.02 \\
    \bottomrule
  \end{tabular*}
\end{table*}

\subsection{Point-Based Downstream Probe with PTv3}
\label{sec:ptv3_probe}

To test whether STEP-Parts also provides stronger supervision for dataset-level learning, we conduct a controlled downstream experiment using PTv3~\citep{Wu2023PointTV}. We construct a paired split of 2{,}500 CAD models, divided into 2{,}000 training shapes, 300 validation shapes, and 200 held-out test shapes. For each model, we use the same STEP-derived tessellated carrier for both supervision settings, so the input geometry is identical across conditions. To avoid conflating supervision quality with the visible-label ceiling of the released PartField exporter, this probe is restricted to shapes with at most 20 STEP parts.

For each shape, we sample $2\times10^4$ points and triangle normals on the common carrier. STEP labels are read directly from the STEP-derived carrier. PartField supervision requires an additional alignment and label-transfer step: the PartField mesh is aligned to STEP space, and PF labels are transferred to the sampled STEP points by nearest-neighbor lookup on dense PF surface samples. PTv3 produces $\ell_2$-normalised 128-dimensional point embeddings trained with a cosine discriminative loss. At inference, point embeddings are clustered with $K$-means using the reference part count on the evaluation carrier (oracle $K$); clusters are then aligned to the reference labels by Hungarian matching. This probe therefore evaluates supervision quality under known part count rather than part-count estimation.

Two models are trained under identical conditions and differ only in the supervision source. Table~\ref{tab:ptv3_probe} reports the held-out test results under STEP-based evaluation. STEP-Parts supervision yields higher performance overall (0.60 vs.\ 0.51 mIoU) and in every complexity bin, with the largest gains on the more complex shapes. Since both models observe the same carrier geometry and differ only in the label source, this improvement is attributable to supervision quality rather than privileged geometric features. 

\begin{table}[t]
\centering
\caption{PTv3 downstream probe on a held-out test split of 200 ABC shapes. Both models share the same PTv3 backbone, embedding head, and training protocol; only the supervision source differs. Evaluation is against the STEP-based reference using oracle part count $K$ and Hungarian label alignment. Best values are shown in \textbf{bold}.}
\label{tab:ptv3_probe}
\small
\begin{tabular}{lccccc}
\toprule
Supervision & mIoU$\uparrow$ & Acc.$\uparrow$ & K=2--4 & K=5--9 & K=10--19 \\
\midrule
PartField  & 0.51 & 0.68 & 0.66 & 0.45 & 0.39 \\
STEP-Parts & \textbf{0.60} & \textbf{0.76} & \textbf{0.71} & \textbf{0.58} & \textbf{0.51} \\
\midrule
$\Delta$   & +0.09 & +0.08 & +0.05 & +0.13 & +0.12 \\
\bottomrule
\end{tabular}
\end{table}

\begin{table}[t]
\centering
\small
\setlength{\tabcolsep}{4pt}
\caption{Cross-evaluation of the PTv3 downstream probe on the held-out test split (200 shapes). Both models use the same carrier geometry, architecture, and optimization settings and differ only in the supervision source. Evaluation against STEP remains the primary reference; PF evaluation is reported as a secondary consistency check on the same carrier. GTc denotes a cluster-free reference-center assignment diagnostic.}
\label{tab:ptv3_cross_eval}
\begin{tabular}{lccc|ccc}
\toprule
& \multicolumn{3}{c|}{Eval on STEP} & \multicolumn{3}{c}{Eval on PF} \\
Supervision & mIoU$\uparrow$ & Acc.$\uparrow$ & GTc$\uparrow$ & mIoU$\uparrow$ & Acc.$\uparrow$ & GTc$\uparrow$ \\
\midrule
PartField  & 0.51 & 0.68 & 0.68 & 0.52 & 0.69 & 0.66 \\
STEP-Parts & \textbf{0.60} & \textbf{0.76} & \textbf{0.79} & \textbf{0.57} & \textbf{0.72} & \textbf{0.69} \\
\bottomrule
\end{tabular}
\end{table}

Cross-evaluation on the same held-out carrier (Table~\ref{tab:ptv3_cross_eval}) preserves the ranking: the model trained with STEP-Parts supervision remains stronger under PF evaluation as well. PF labels are not treated as ground truth in this paper; rather, this secondary check indicates that the observed advantage is not merely an artifact of matching the STEP reference.

\begin{table}[t]
\centering
\small
\setlength{\tabcolsep}{5pt}
\caption{Test-set PTv3 probe mIoU stratified by STEP part-count complexity.}
\label{tab:supp_ptv3_cross_eval_bins}
\begin{tabular}{lcccc}
\toprule
Supervision / Eval & Overall & $K{=}2$--$4$ & $K{=}5$--$9$ & $K{=}10$--$19$ \\
\midrule
PartField / STEP  & 0.51 & 0.66 & 0.45 & 0.39 \\
STEP-Parts / STEP & \textbf{0.60} & \textbf{0.71} & \textbf{0.58} & \textbf{0.51} \\
PartField / PF    & 0.52 & 0.68 & 0.49 & 0.39 \\
STEP-Parts / PF   & \textbf{0.57} & \textbf{0.68} & \textbf{0.55} & \textbf{0.46} \\
\bottomrule
\end{tabular}
\end{table}

Table~\ref{tab:supp_ptv3_cross_eval_bins} further shows that the benefit of STEP-Parts supervision is concentrated on the more complex shapes. Under STEP evaluation, the mIoU gain increases from 0.05 in the $K{=}2$--4 bin to 0.13 and 0.12 in the $K{=}5$--9 and $K{=}10$--19 bins, respectively. Under PF evaluation, the same pattern remains visible as a secondary consistency check: there is no gain in the simplest bin, whereas the $K{=}5$--9 and $K{=}10$--19 bins improve by 0.06 and 0.07. These stratified results are consistent with the earlier PartField comparison, where disagreement with the STEP reference increases with structural complexity.

\subsection{Scalability and Practicality}
\label{subsec:scalability}

We apply the STEP-Part extraction pipeline to approximately 180{,}000 STEP models from the DeepCAD subset of the ABC dataset using a deterministic CPU-based implementation without GPU acceleration. The processed models span a wide range of geometric complexity, including shapes containing hundreds or thousands of analytic regions.

Figure~\ref{fig:step_many_parts} shows representative examples of models with very high STEP-Part counts. The ability to robustly extract such decompositions demonstrates that the construction is practical at scale and applicable beyond curated or low-complexity benchmarks. Figure~\ref{fig:teaser} shows a selected subset of rendered shapes.

\begin{figure*}[!ht]
  \centering
  \setlength{\tabcolsep}{0pt}%
  \begin{tabular}{@{}ccc@{}}
    \begin{subfigure}[t]{0.32\textwidth}\centering\includegraphics[width=\linewidth]{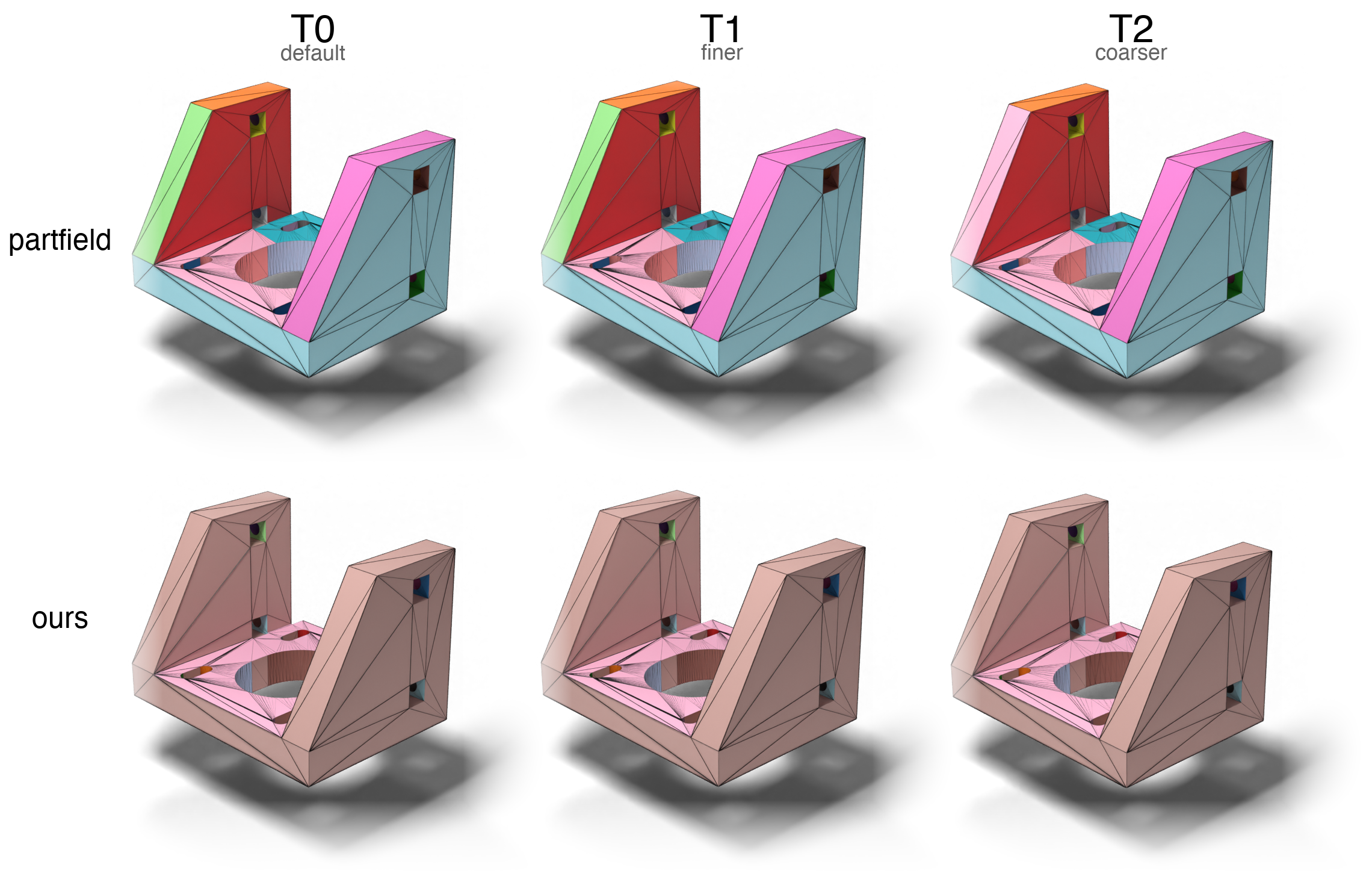}\subcaption*{}\end{subfigure}    &
    \begin{subfigure}[t]{0.32\textwidth}\centering\includegraphics[width=\linewidth]{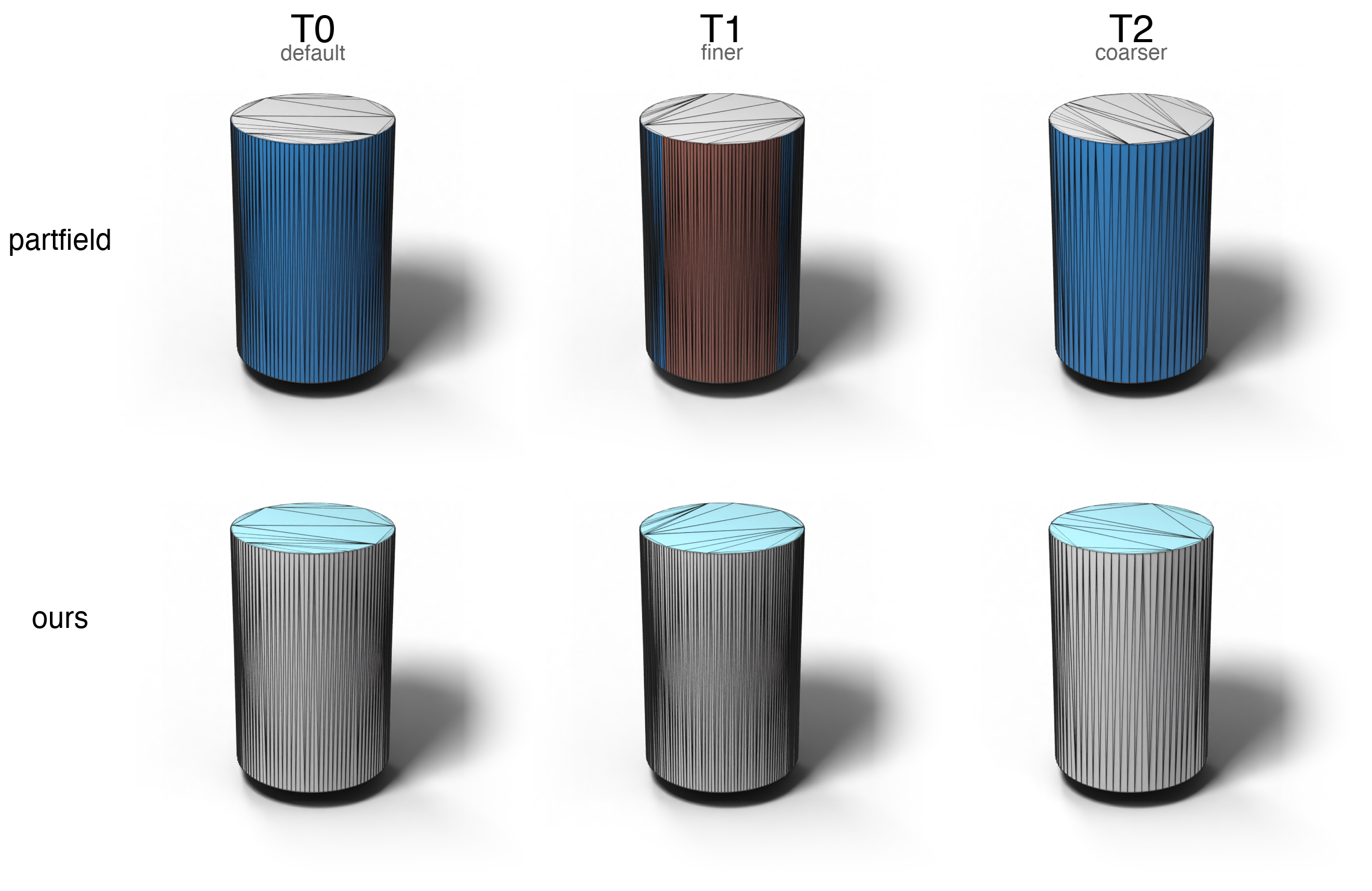}\subcaption*{}\end{subfigure}    &
    \begin{subfigure}[t]{0.32\textwidth}\centering\includegraphics[width=\linewidth]{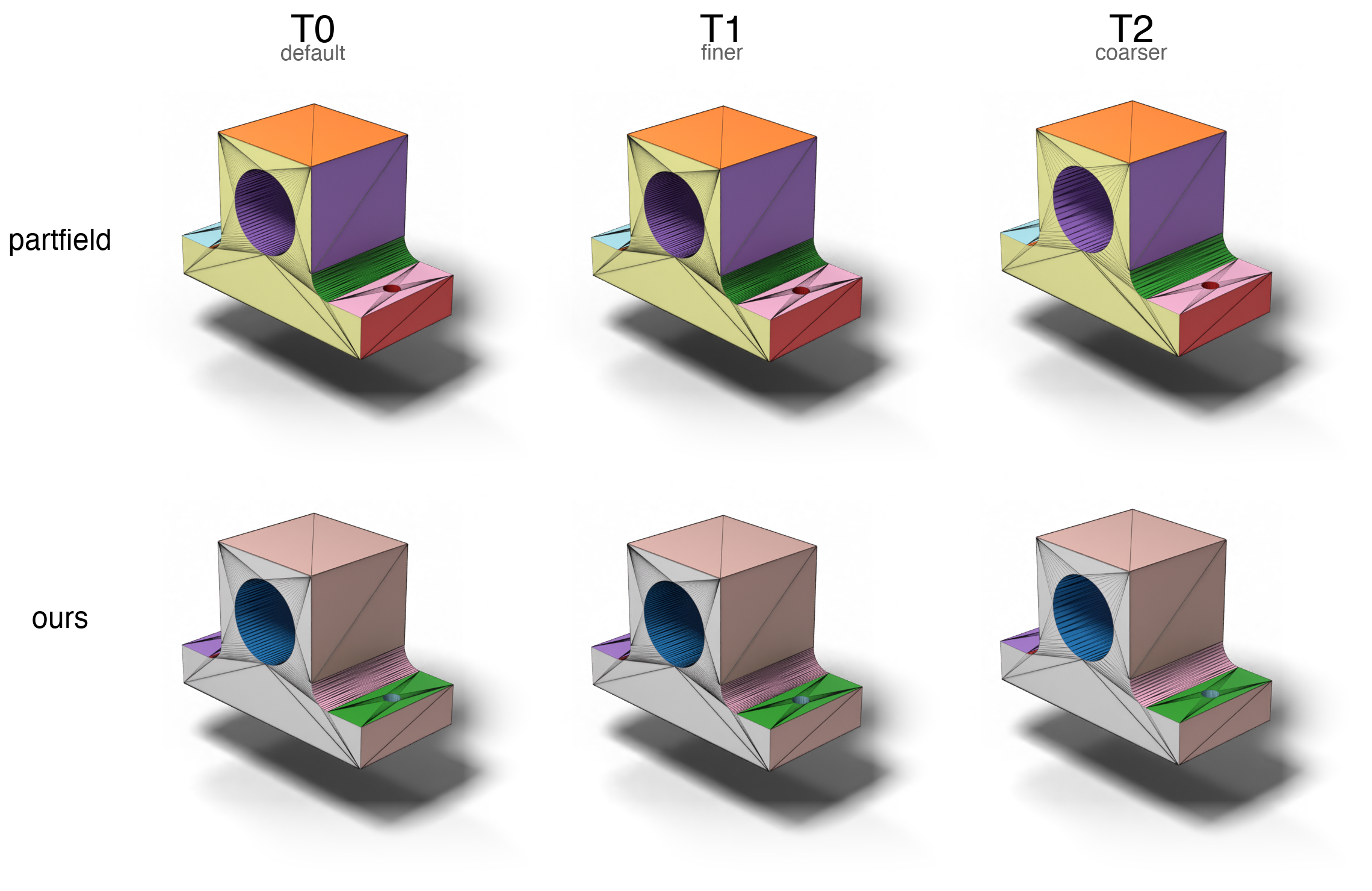}\subcaption*{}\end{subfigure}    \\

    \begin{subfigure}[t]{0.32\textwidth}\centering\includegraphics[width=\linewidth]{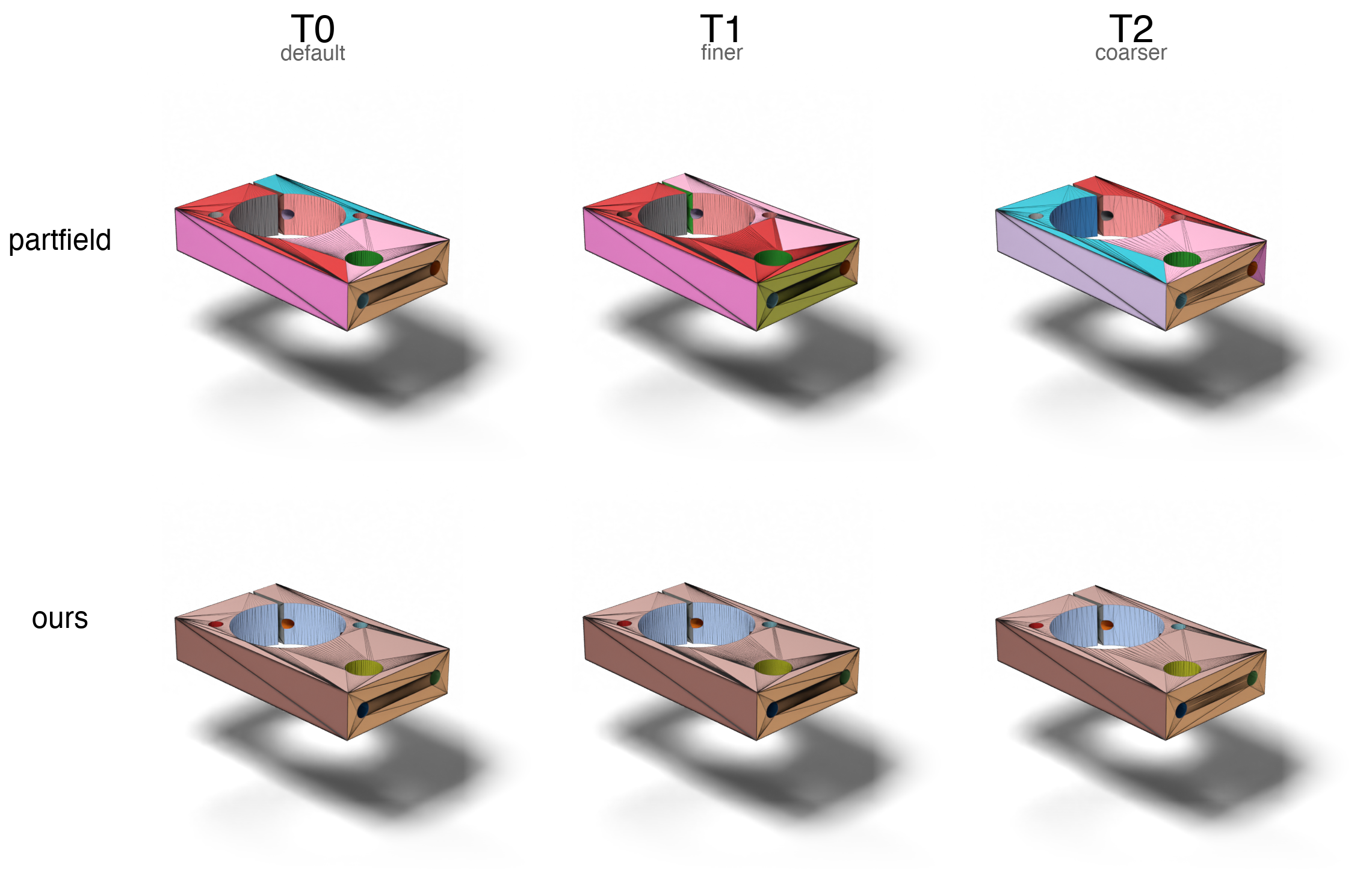}\subcaption*{}\end{subfigure}    &
    \begin{subfigure}[t]{0.32\textwidth}\centering\includegraphics[width=\linewidth]{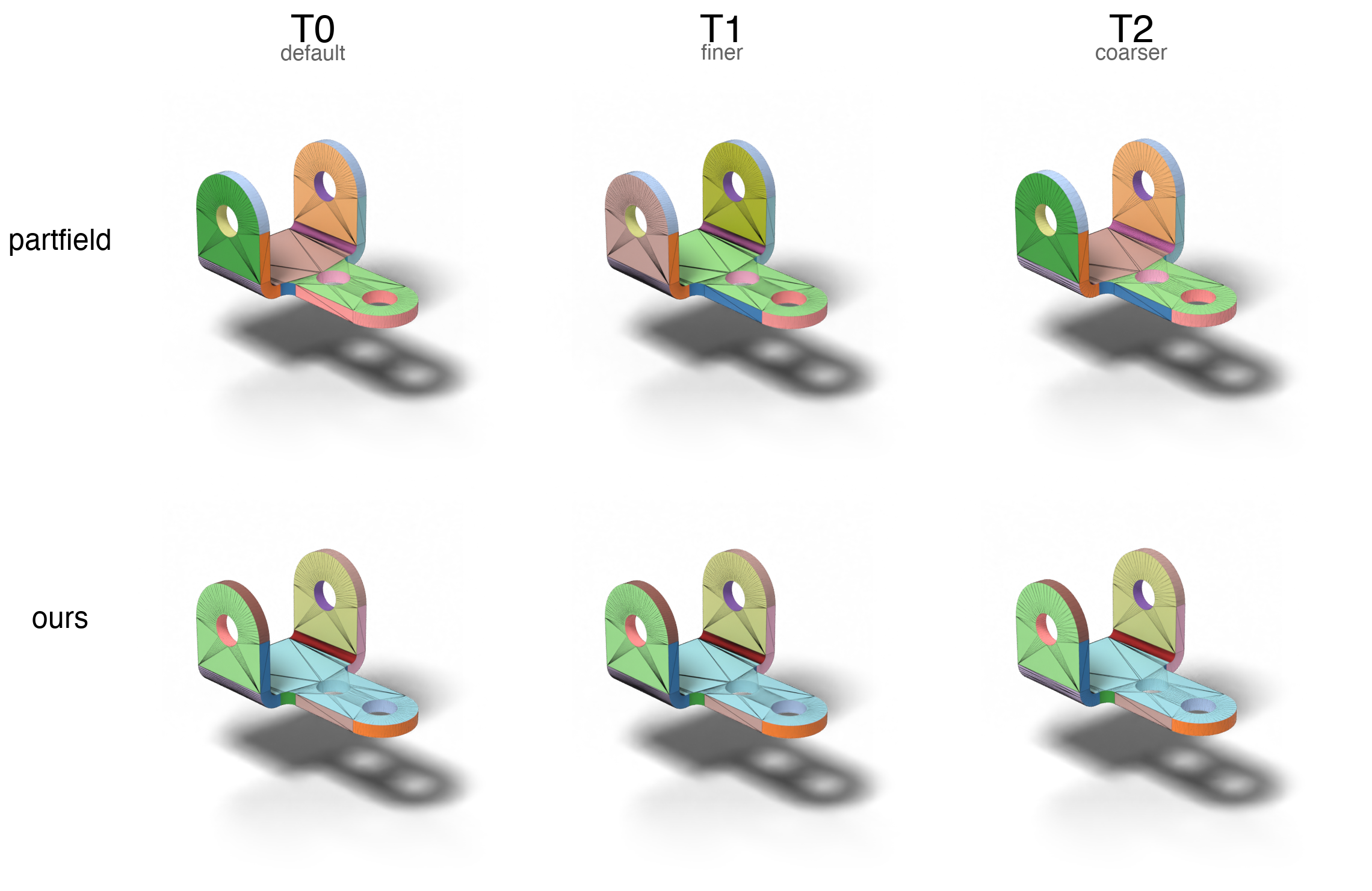}\subcaption*{}\end{subfigure}    &
    \begin{subfigure}[t]{0.32\textwidth}\centering\includegraphics[width=\linewidth]{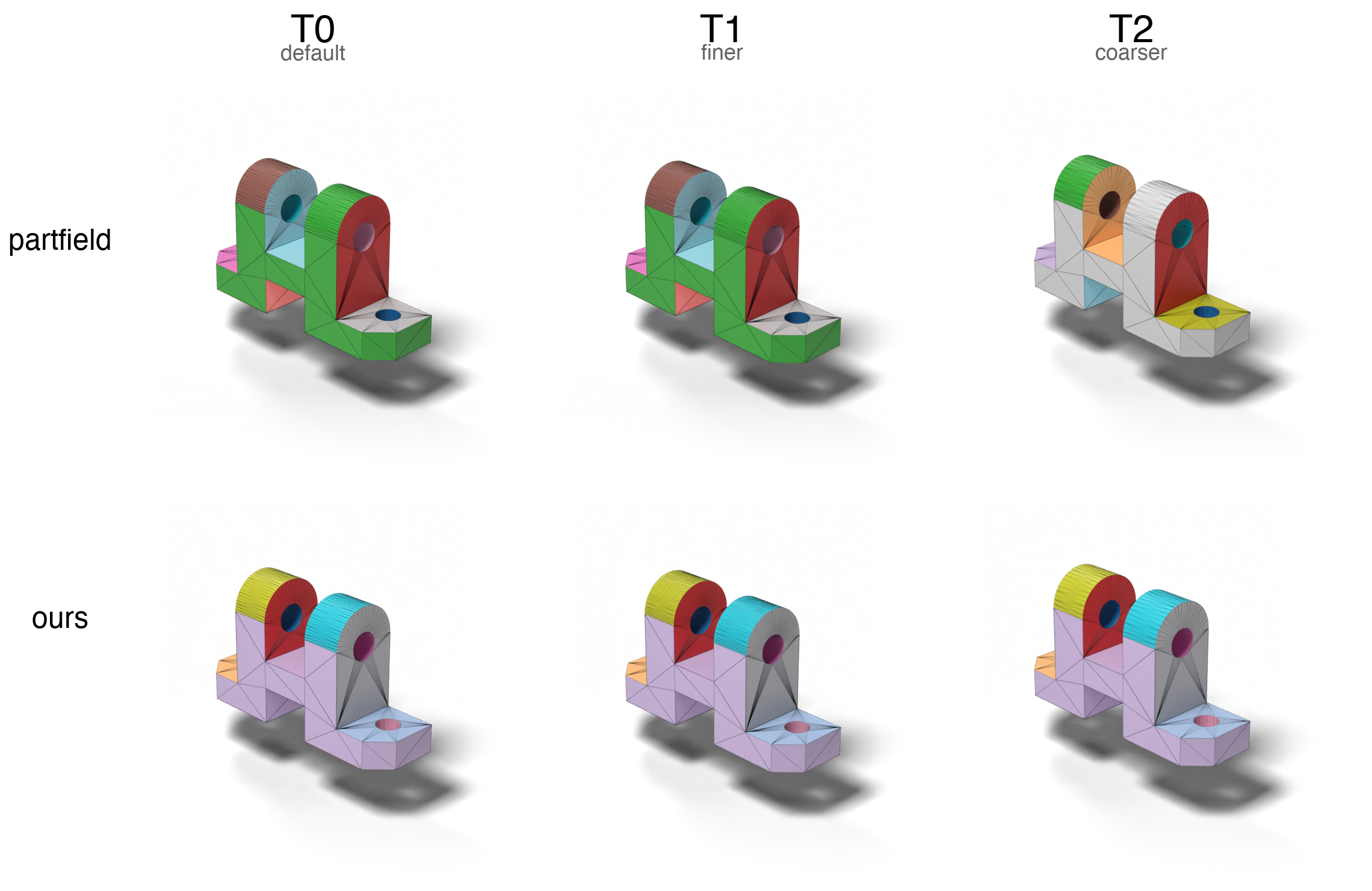}\subcaption*{}\end{subfigure}    \\

    \begin{subfigure}[t]{0.32\textwidth}\centering\includegraphics[width=\linewidth]{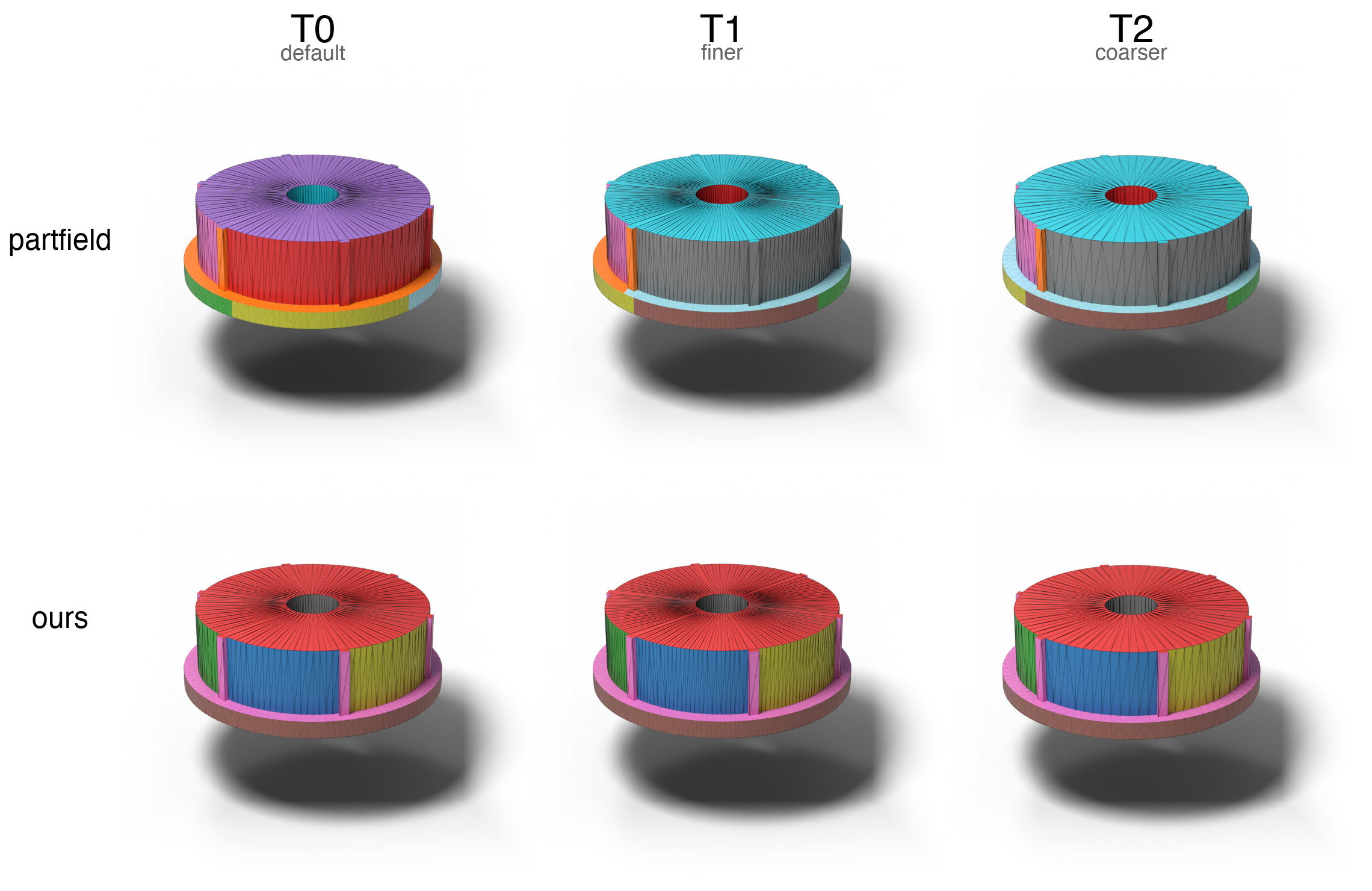}\subcaption*{}\end{subfigure}    &
    \begin{subfigure}[t]{0.32\textwidth}\centering\includegraphics[width=\linewidth]{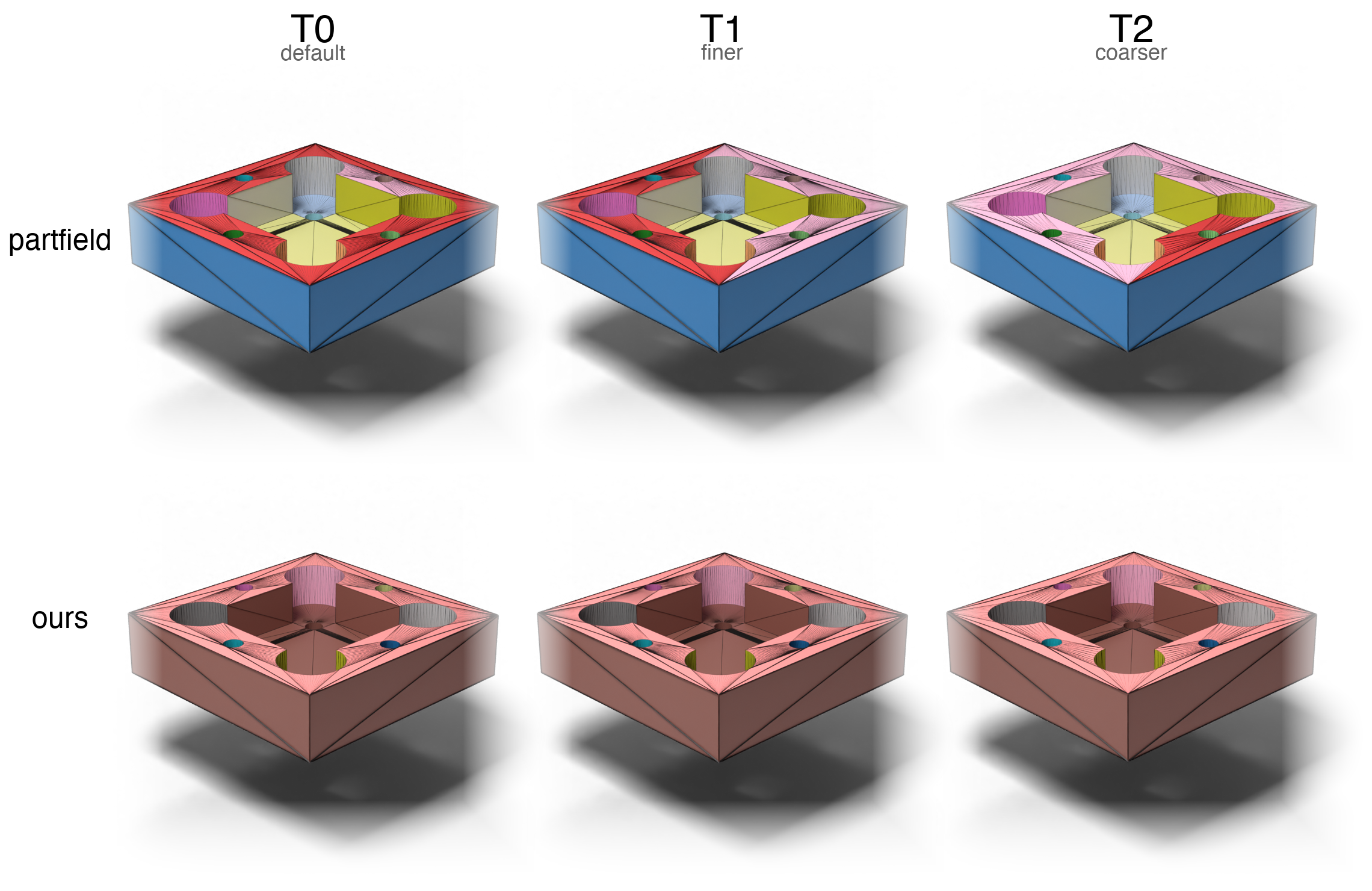}\subcaption*{}\end{subfigure}    &
    \begin{subfigure}[t]{0.32\textwidth}\centering\includegraphics[width=\linewidth]{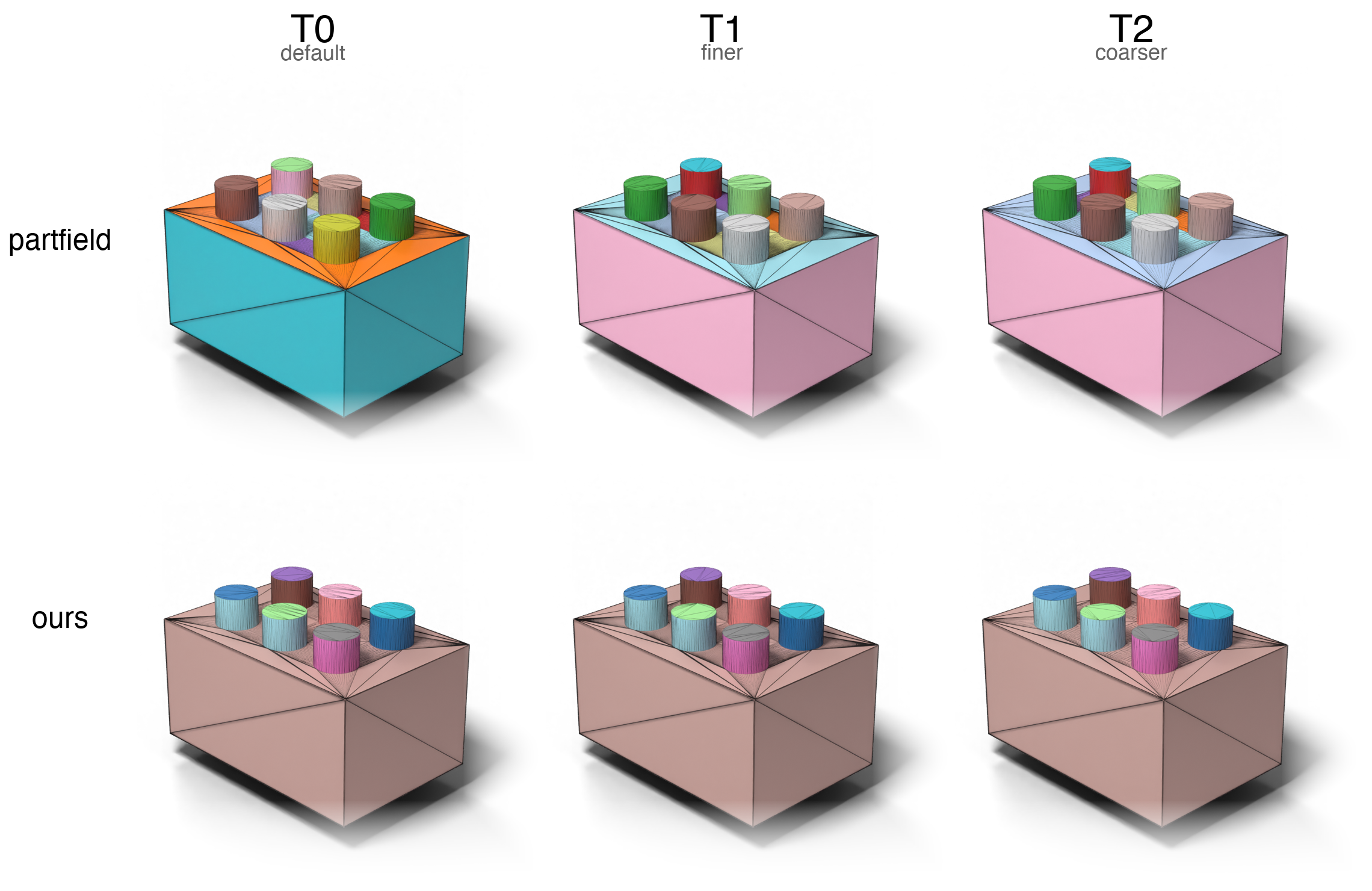}\subcaption*{}\end{subfigure}    \\

    \begin{subfigure}[t]{0.32\textwidth}\centering\includegraphics[width=\linewidth]{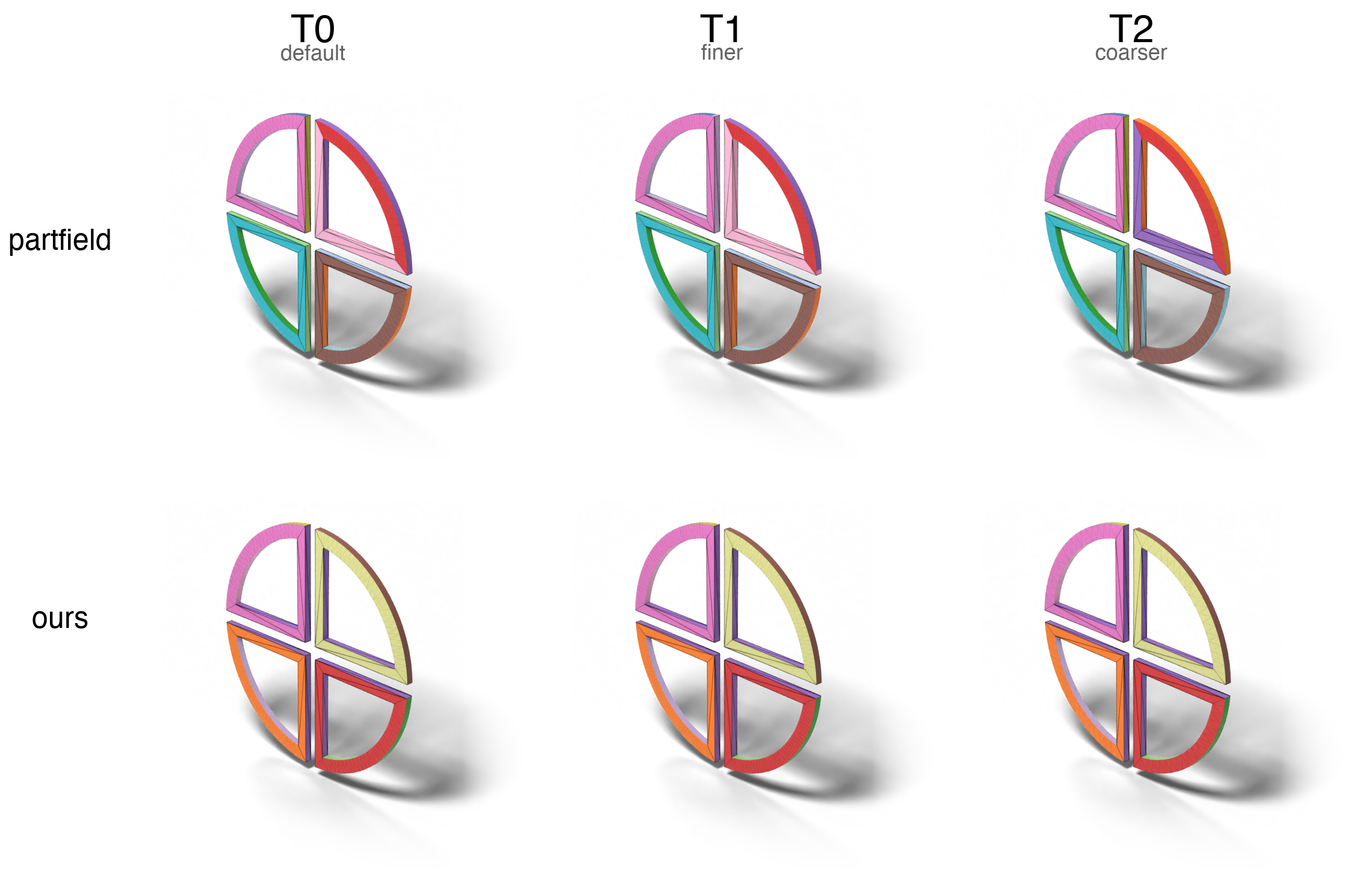}\subcaption*{}\end{subfigure}    &
    \begin{subfigure}[t]{0.32\textwidth}\centering\includegraphics[width=\linewidth]{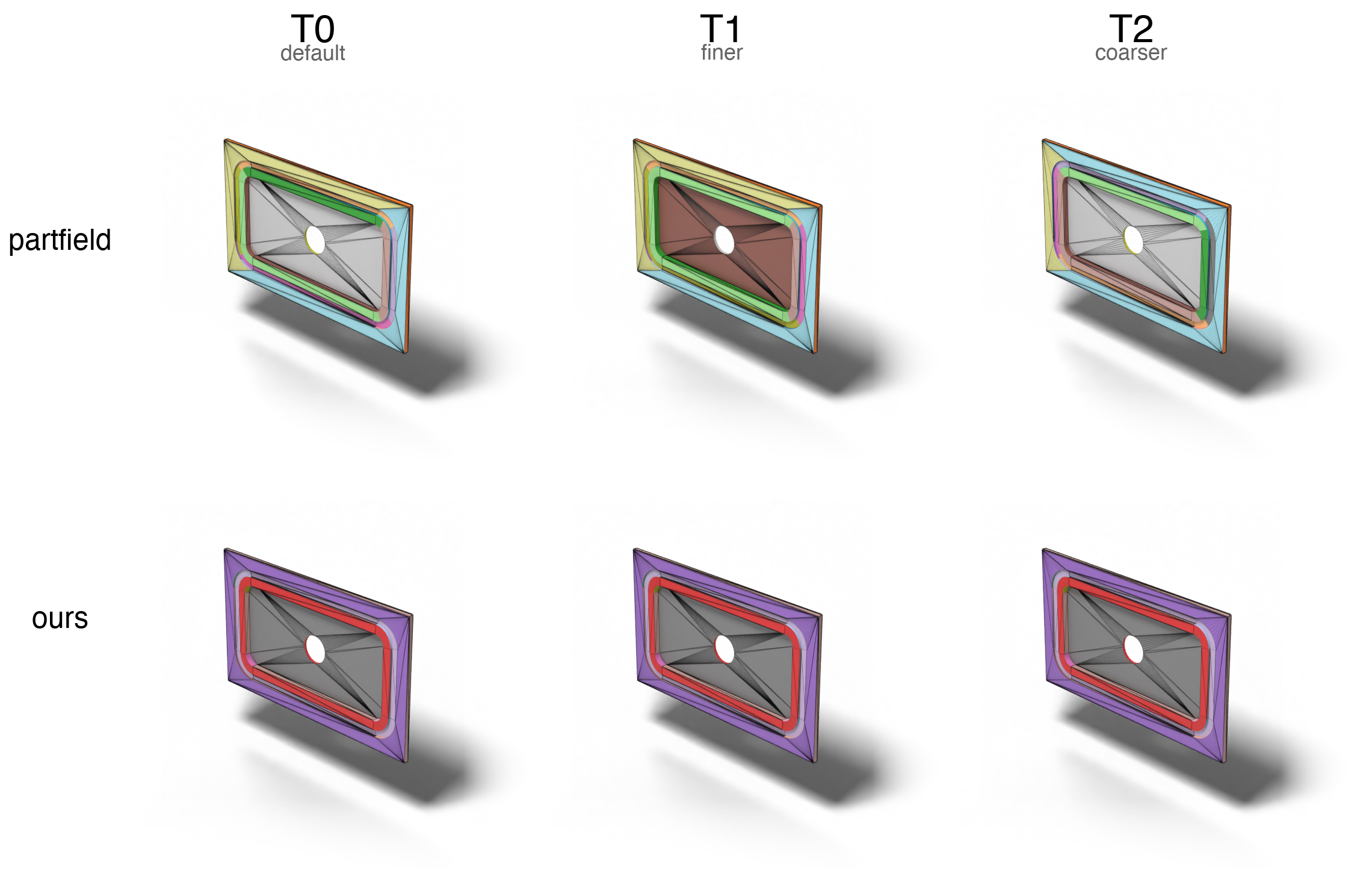}\subcaption*{}\end{subfigure}    &
    \begin{subfigure}[t]{0.32\textwidth}\centering\includegraphics[width=\linewidth]{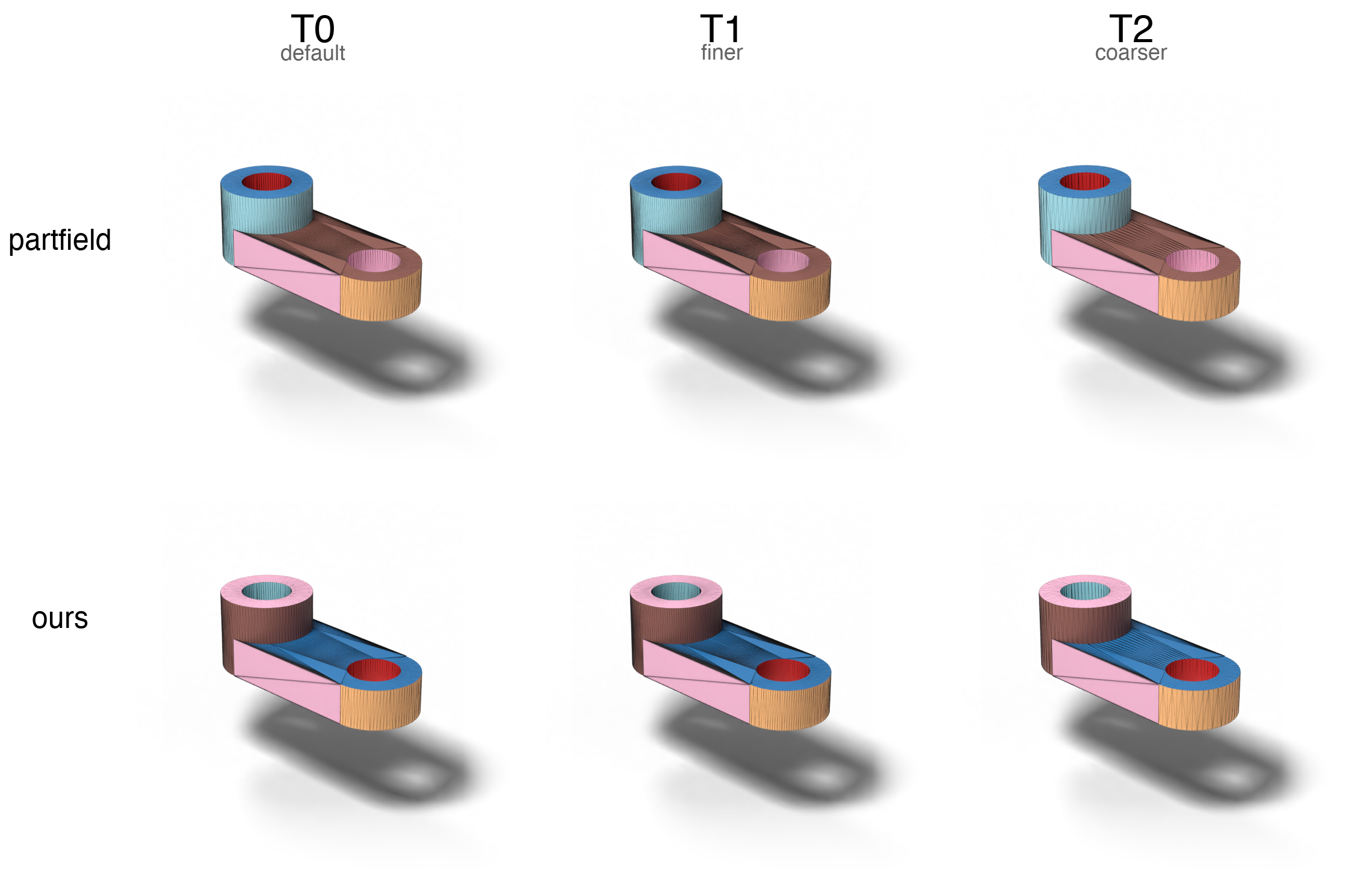}\subcaption*{}\end{subfigure}    \\

    \begin{subfigure}[t]{0.32\textwidth}\centering\includegraphics[width=\linewidth]{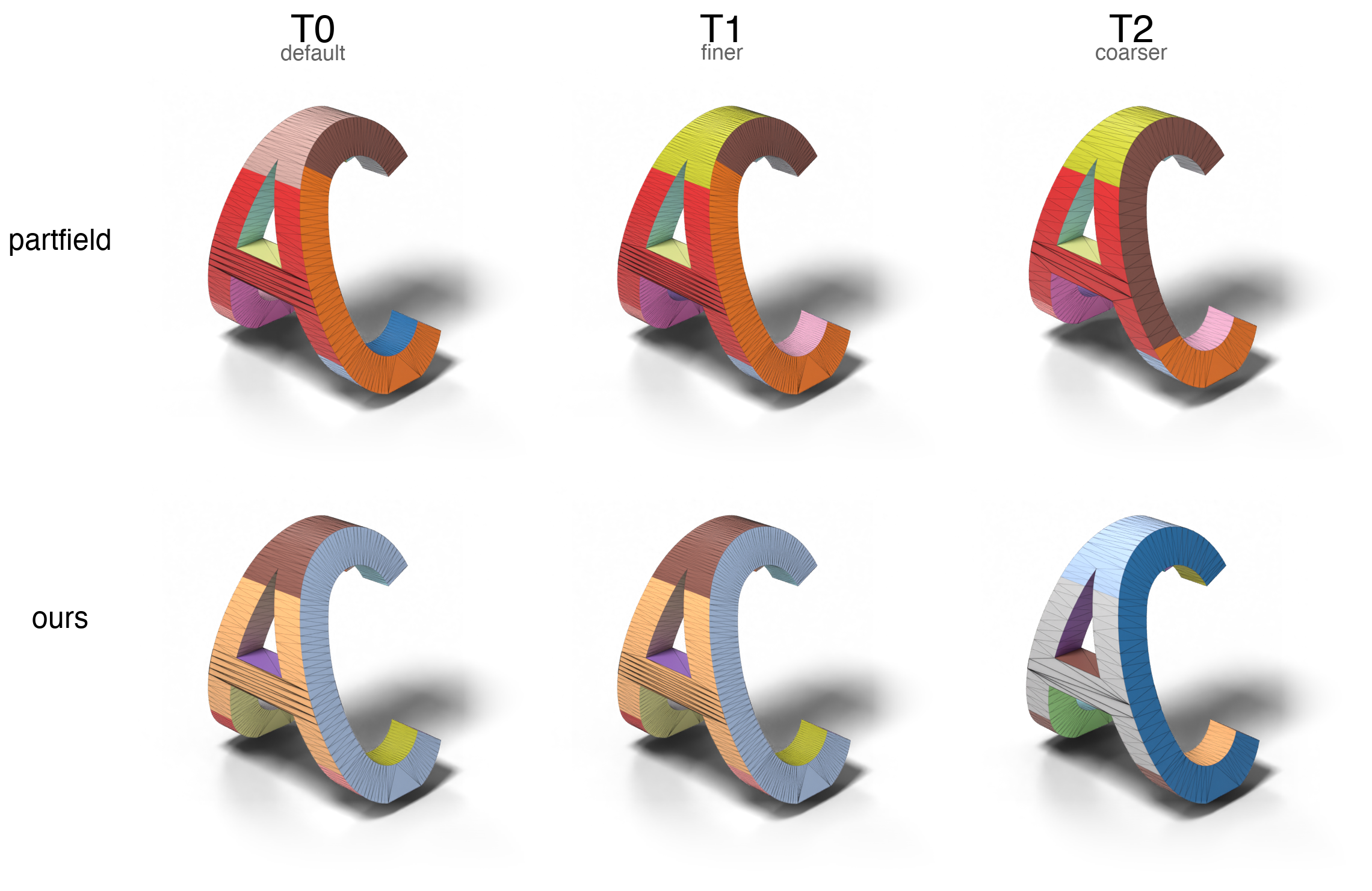}\subcaption*{}\end{subfigure}    &
    \begin{subfigure}[t]{0.32\textwidth}\centering\includegraphics[width=\linewidth]{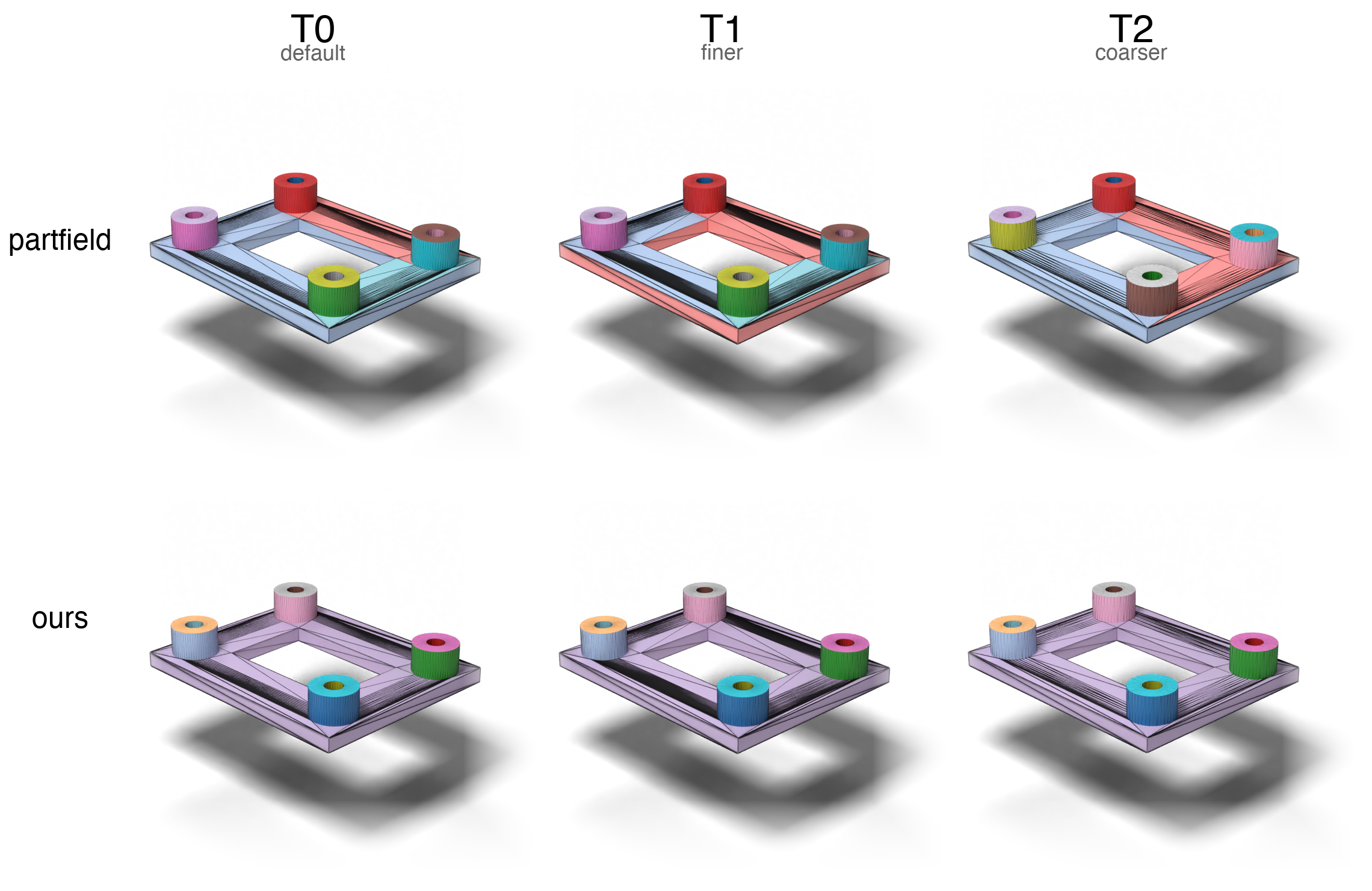}\subcaption*{}\end{subfigure}    &
    \begin{subfigure}[t]{0.32\textwidth}\centering\includegraphics[width=\linewidth]{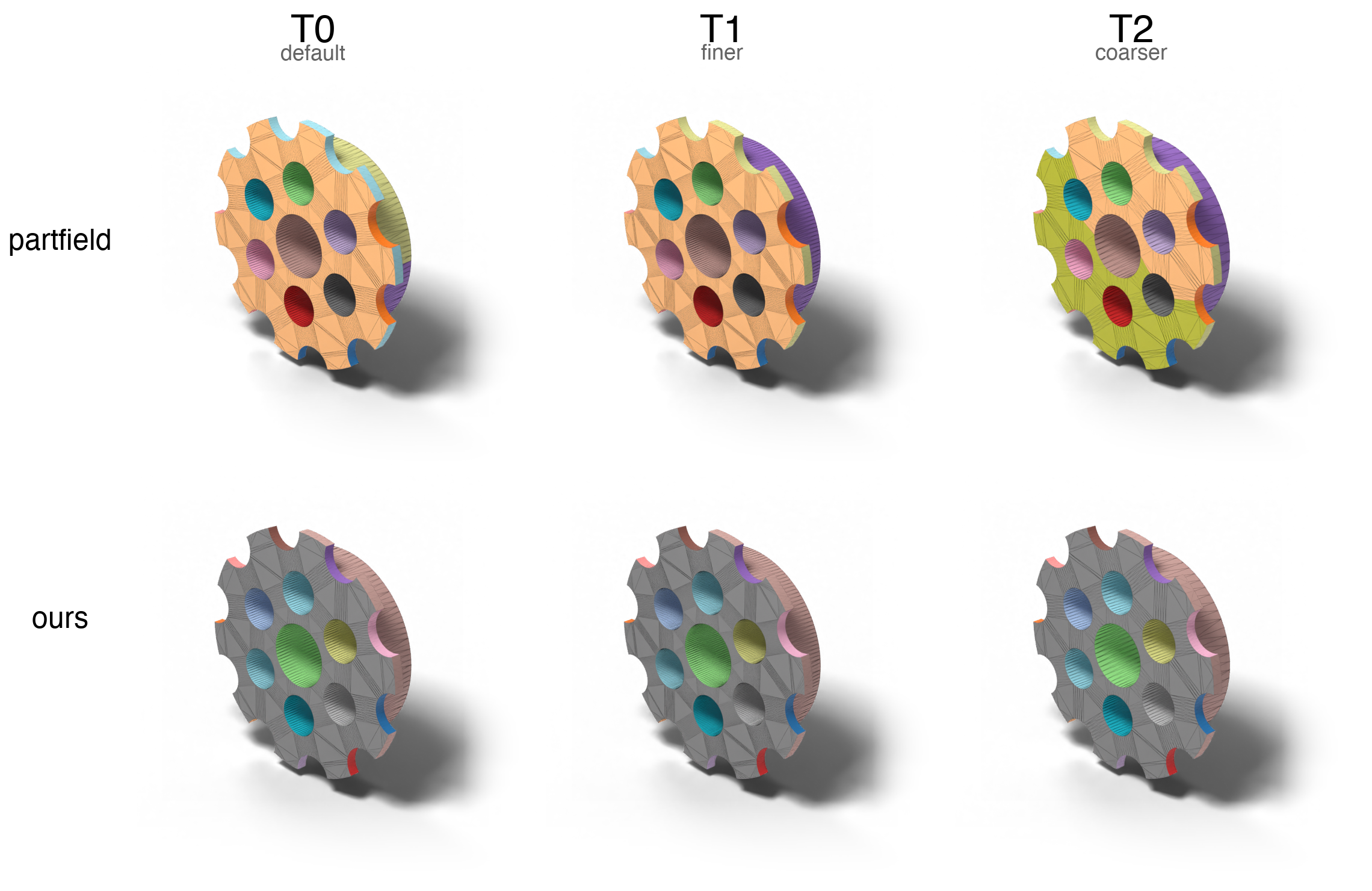}\subcaption*{}\end{subfigure}
  \end{tabular}
  \caption{Comparison of a subset of models on their STEP-partitioning versus PartField segmentation outcomes under different tessellation levels.}
  \label{fig:tess_comp}
\end{figure*}

\section{Limitations and Practical Considerations}

\begin{figure*}[!ht]
  \centering
  \setlength{\tabcolsep}{0pt}
  \begin{tabular}{@{}ccc@{}}
        \includegraphics[width=0.32\textwidth]{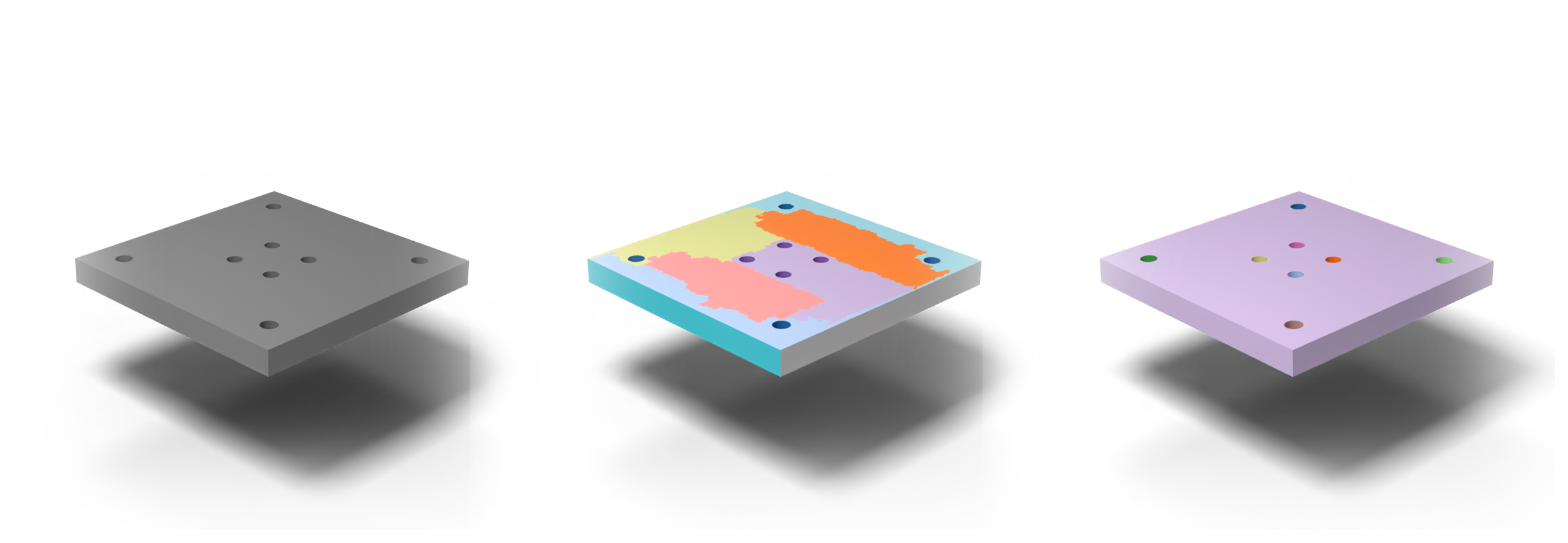} &
        \includegraphics[width=0.32\textwidth]{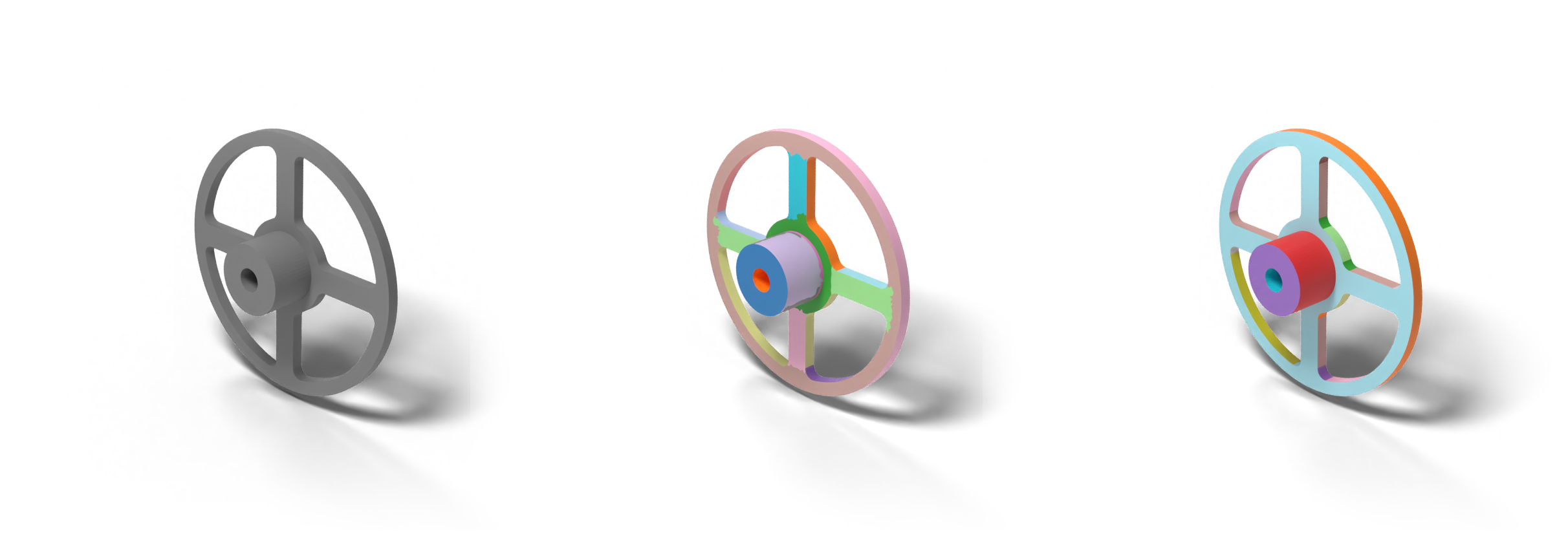} &
        \includegraphics[width=0.32\textwidth]{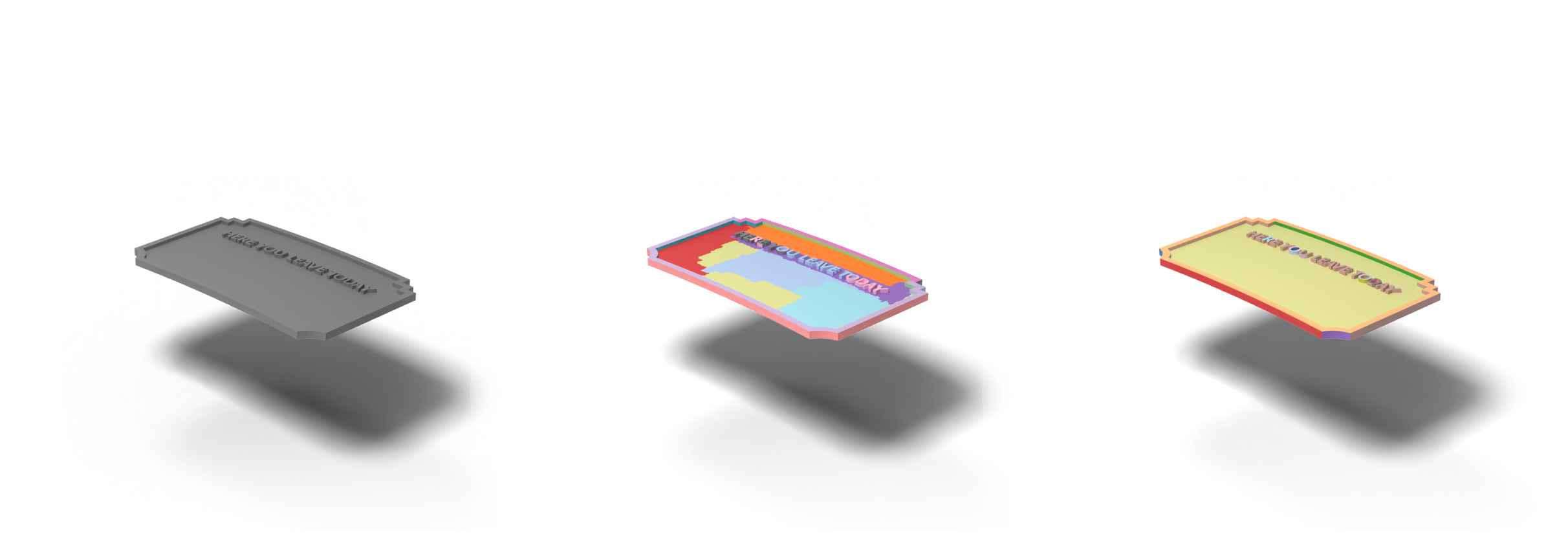} \\
        \includegraphics[width=0.32\textwidth]{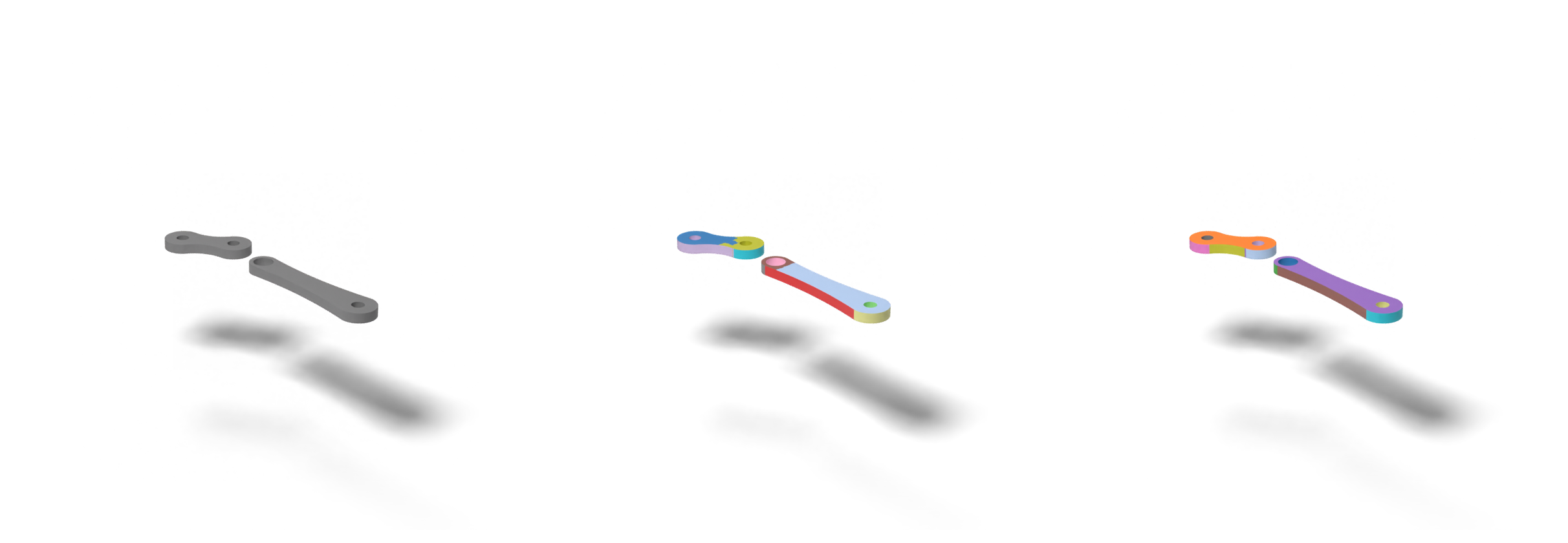} &
        \includegraphics[width=0.32\textwidth]{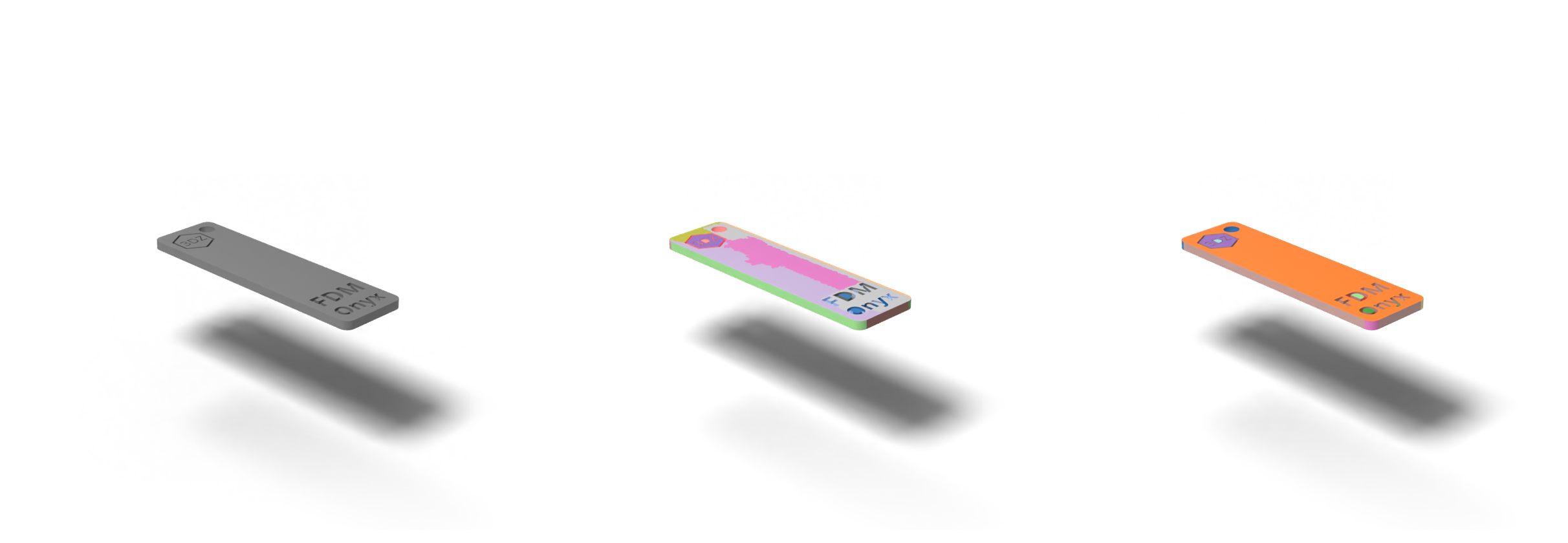} &
        \includegraphics[width=0.32\textwidth]{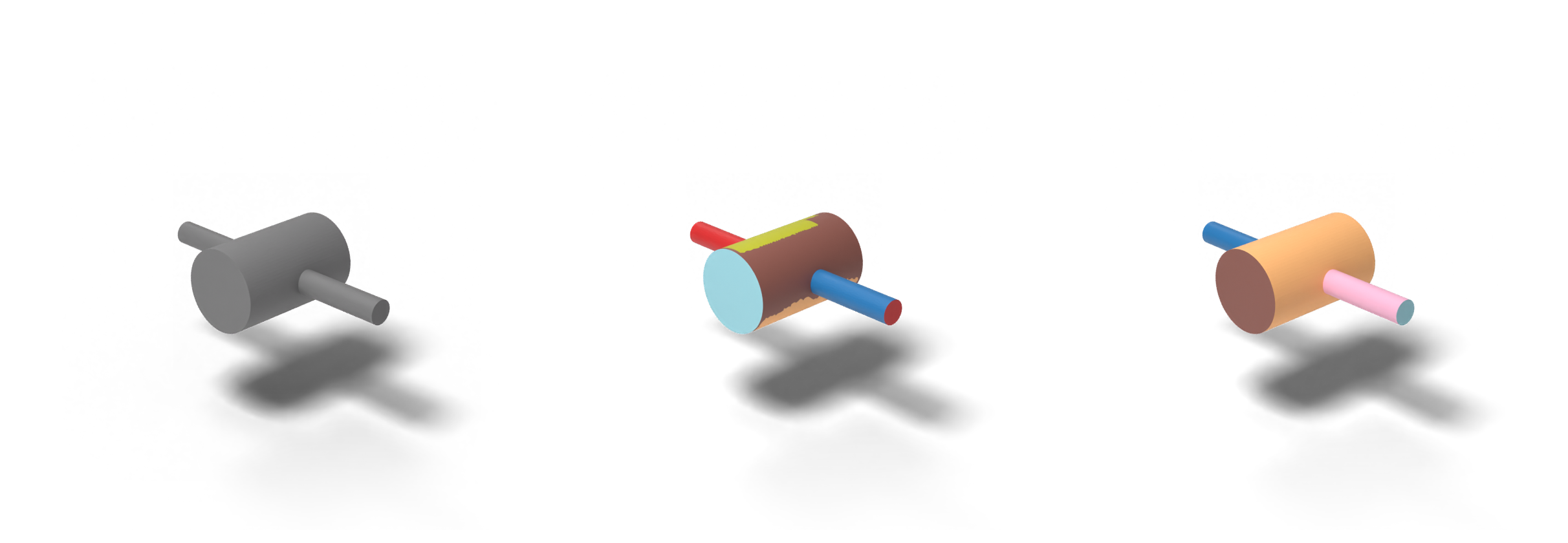} \\
        \includegraphics[width=0.32\textwidth]{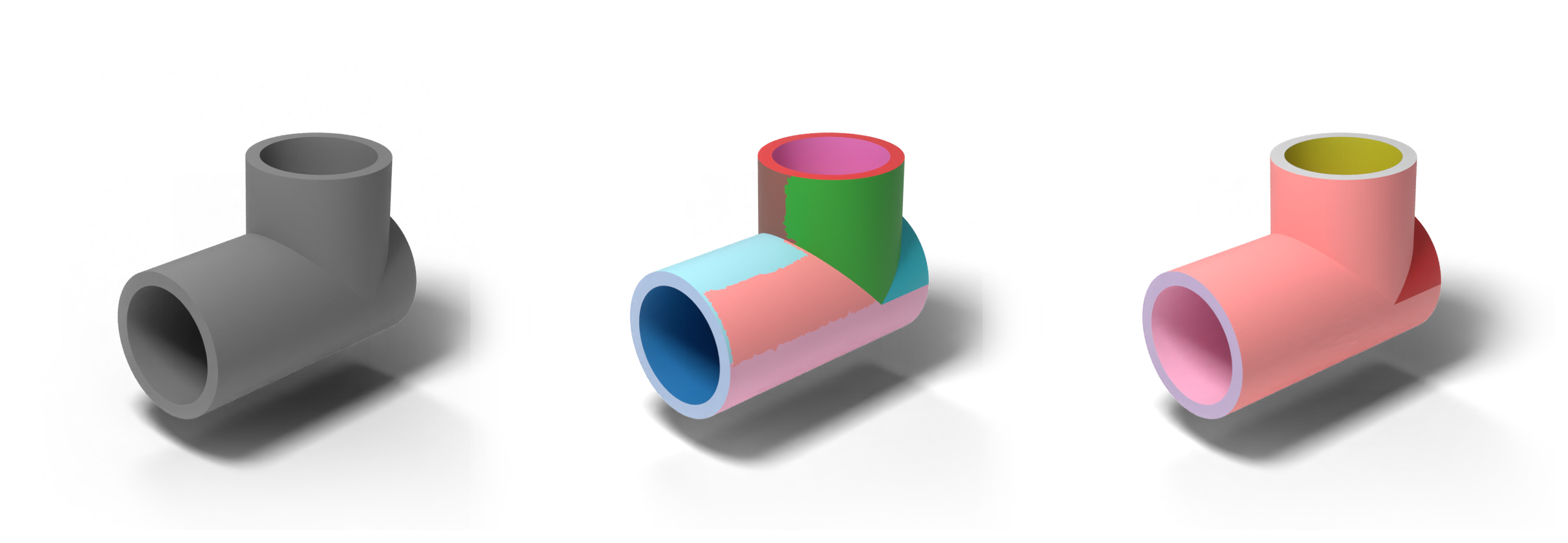} &
        \includegraphics[width=0.32\textwidth]{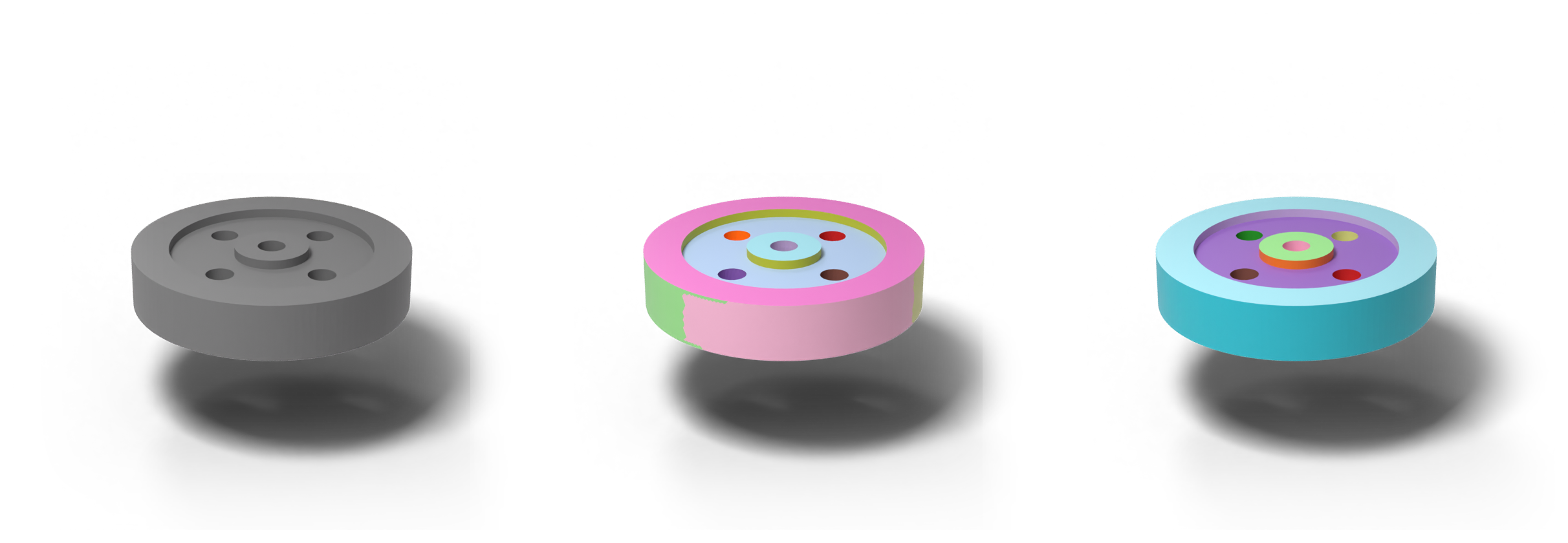} &
        \includegraphics[width=0.32\textwidth]{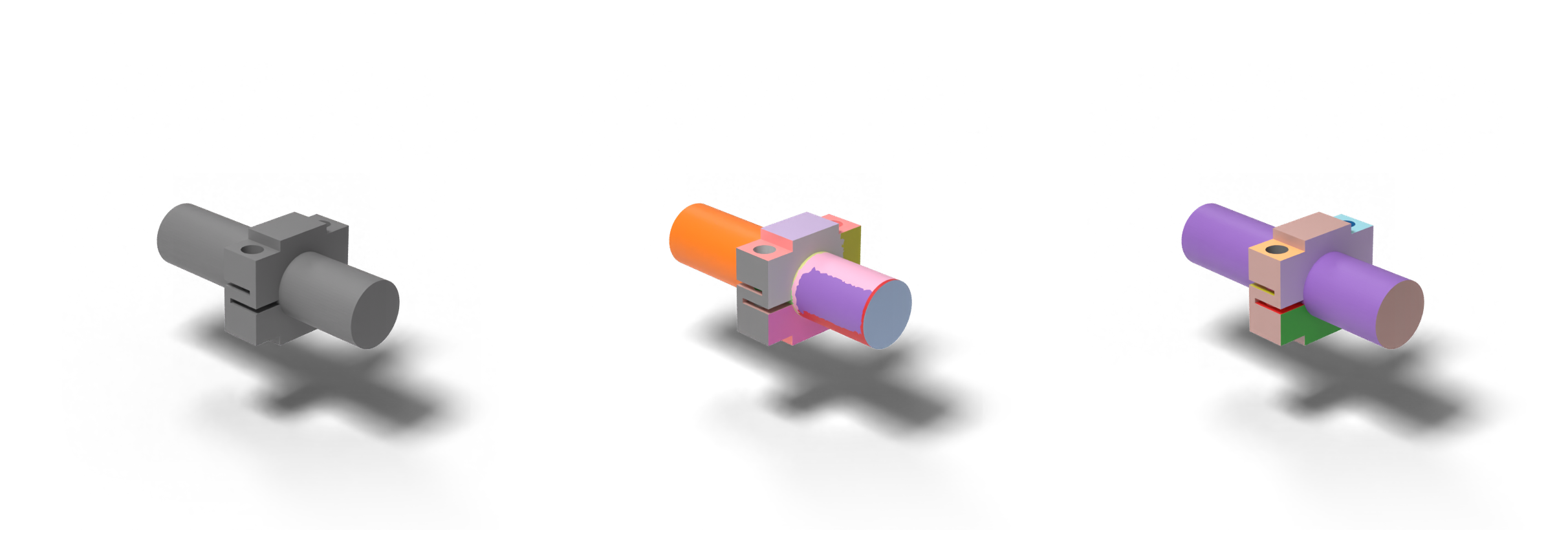} \\
        \includegraphics[width=0.32\textwidth]{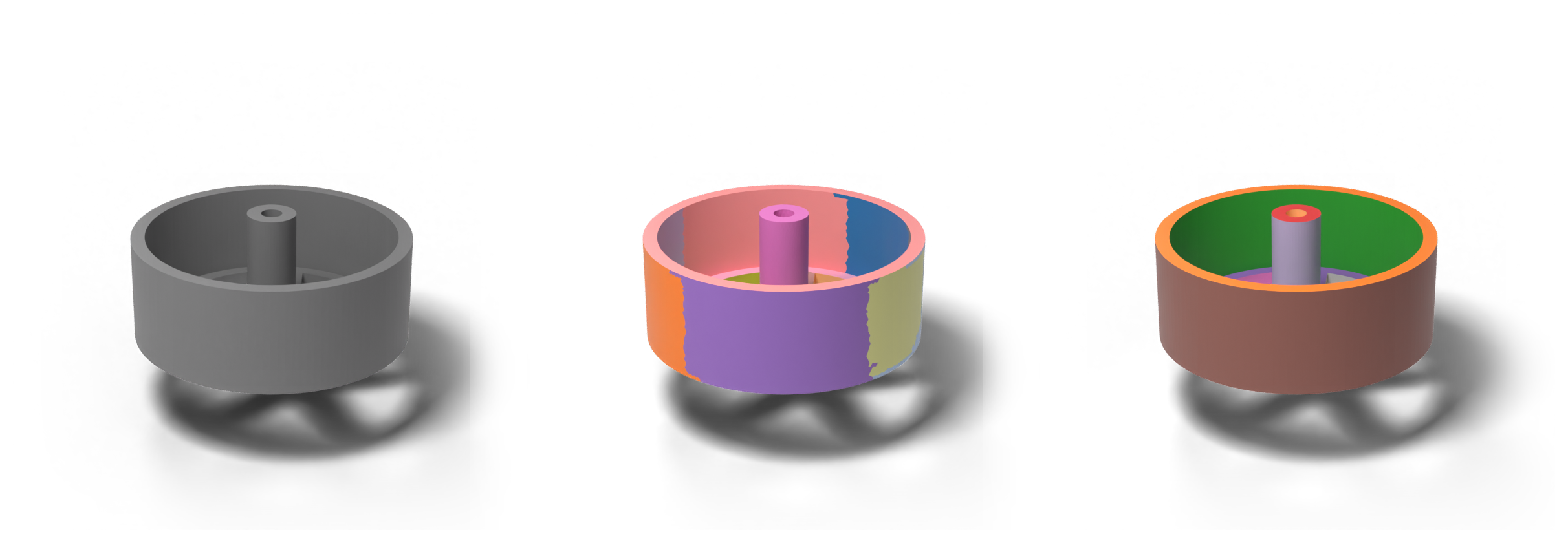} &
        \includegraphics[width=0.32\textwidth]{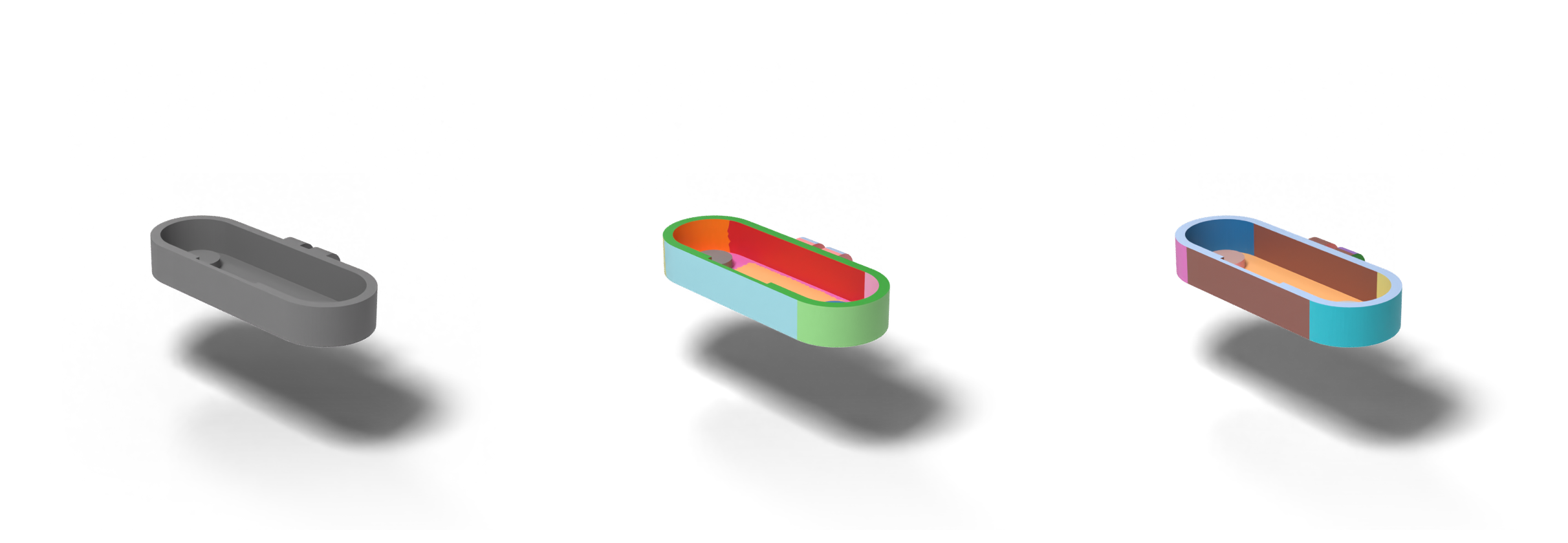} &
        \includegraphics[width=0.32\textwidth]{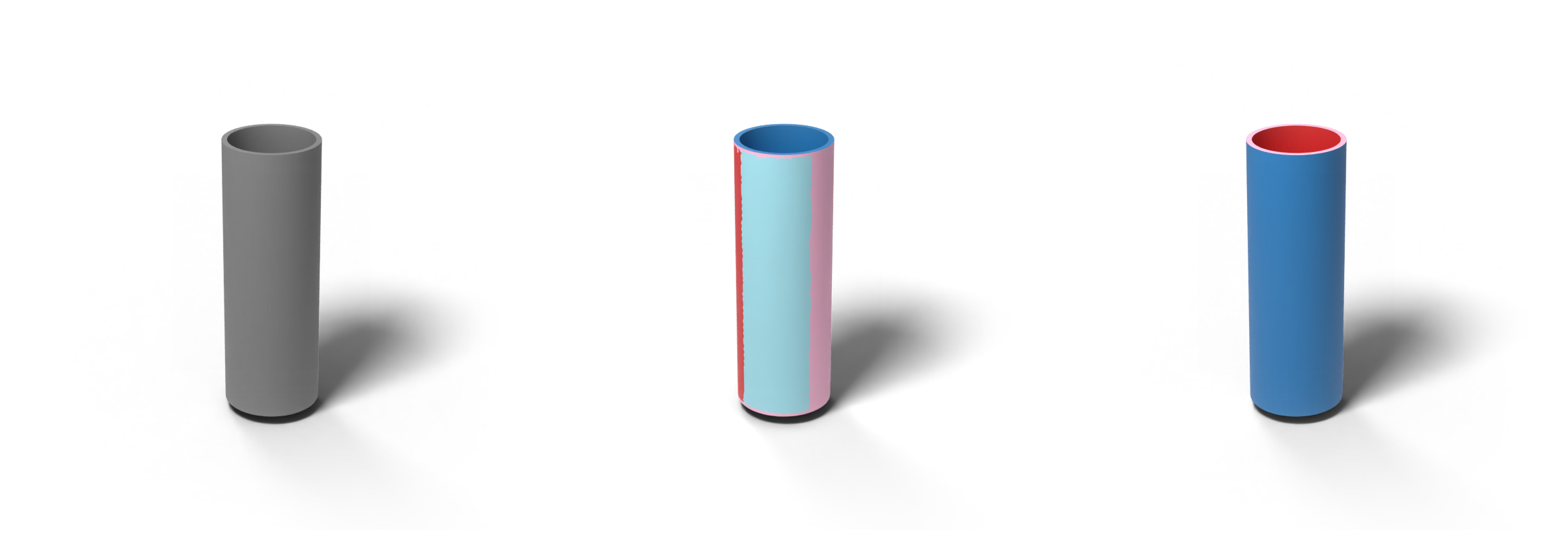} \\
        \includegraphics[width=0.32\textwidth]{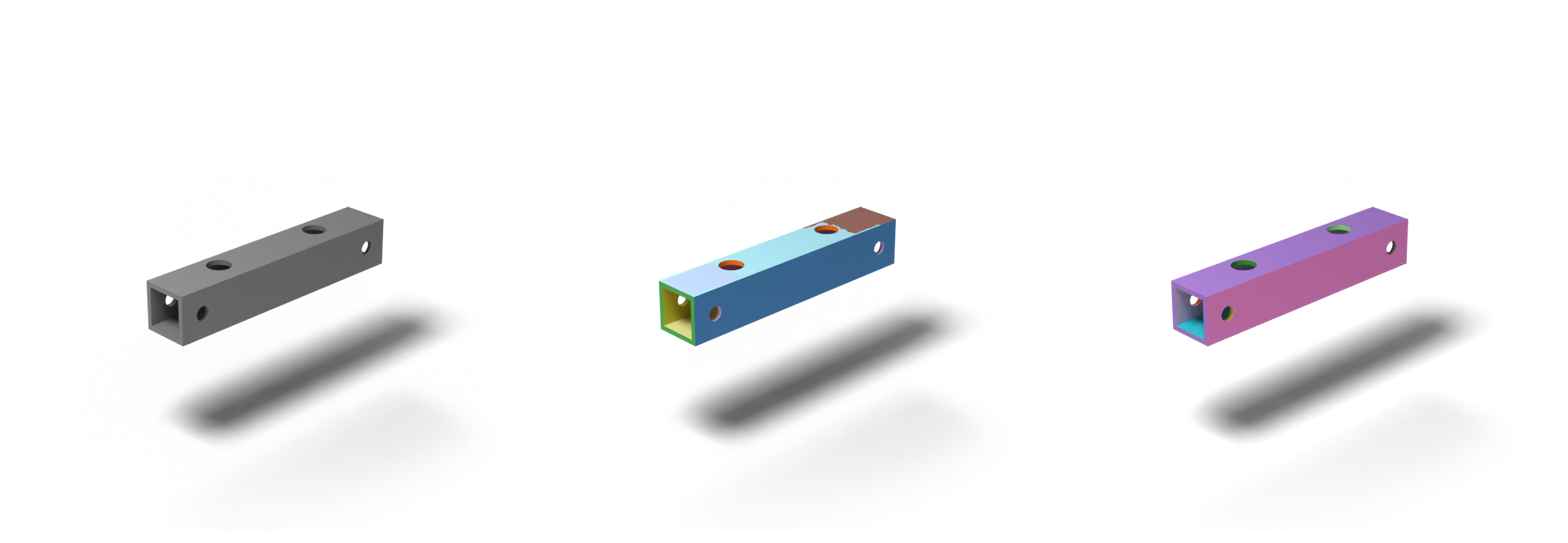} &
        \includegraphics[width=0.32\textwidth]{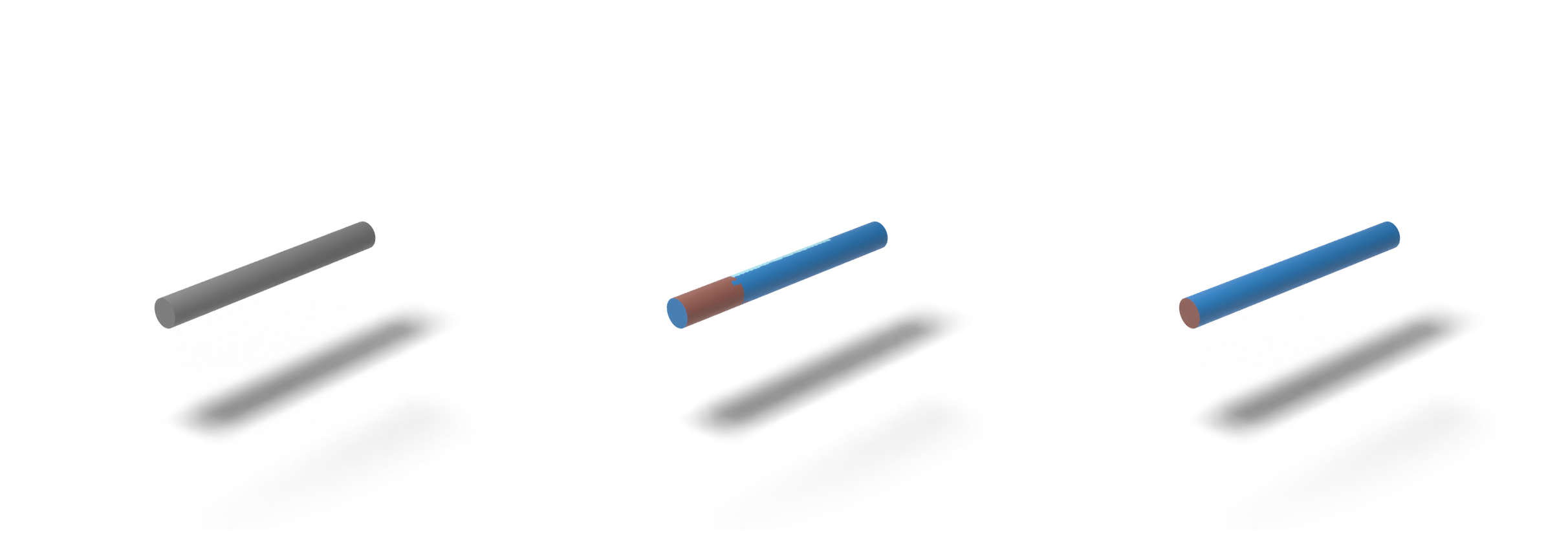} &
        \includegraphics[width=0.32\textwidth]{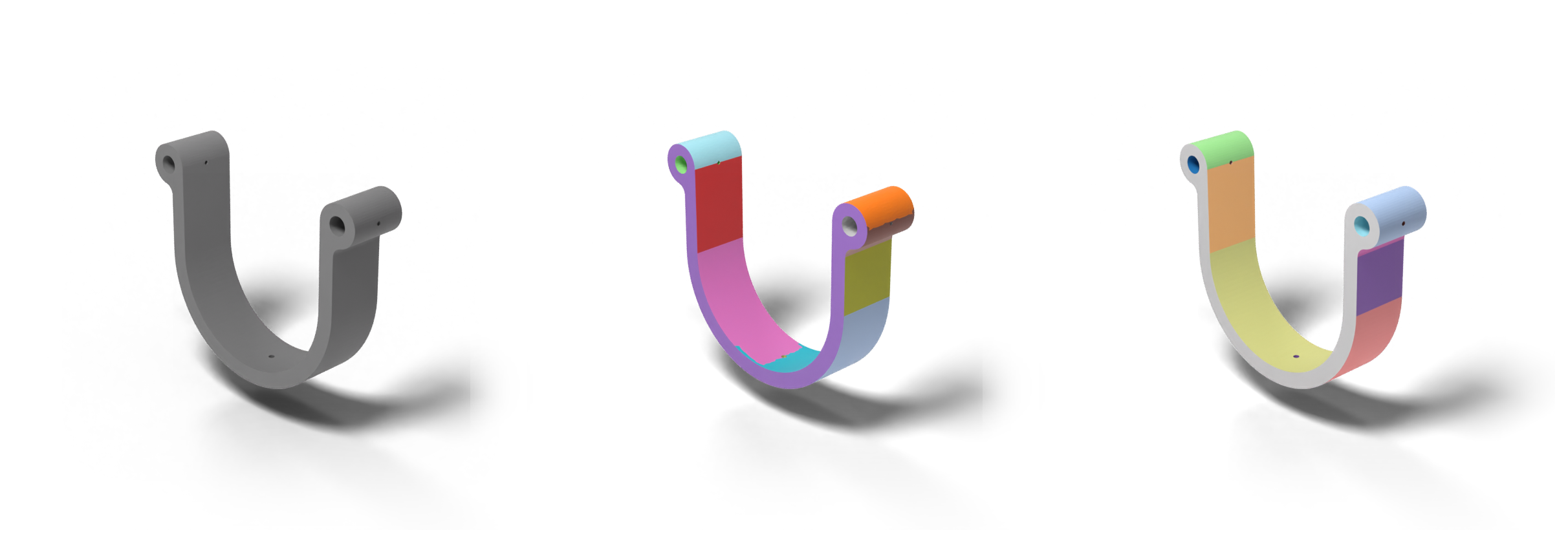} \\
  \end{tabular}
  \caption{Comparison of a subset of models on their STEP-partitioning versus PartField segmentation. From left to right: ground truth, PartField, our method}
  \label{fig:results_grid}
\end{figure*}

\paragraph{Geometric scope}
{STEP-Parts} is a topology-faithful geometric decomposition defined by analytic surface compatibility and local continuity. It is not intended to recover semantic or functional parts, and it may disagree with human notions of part boundaries when modeling intent, assembly structure, or functional grouping diverge from analytic primitive structure.

\paragraph{Discretized carriers and tessellation extremes}
While the partition is defined on the B-Rep, all visualizations and many quantitative comparisons are performed after projecting labels onto a tessellated surface. As a result, extremely coarse tessellations can degrade the discrete carrier: thin regions may lose triangle support, and boundary localization can become less faithful on the mesh even when the underlying B-Rep predicate is unchanged. This effect appears primarily in the tails (outlier shapes) of tessellation self-consistency and should be interpreted as a limitation of the discretized representation rather than a change in the intrinsic definition.

\paragraph{Small-component handling}
For robustness of the discrete carrier, we optionally suppress or absorb very small isolated components (e.g., below a minimum triangle-support threshold). This step is a practical hygiene operation to reduce degenerate fragments caused by trimming and tessellation artifacts; however, it can remove or merge extremely thin features on aggressively coarsened meshes. When reporting quantitative results, it should therefore be stated explicitly whether this post-processing is enabled and whether it is applied uniformly across all tessellations.

\paragraph{Threshold selection}
The dihedral threshold $\theta$ is used as a local continuity discriminator within a low-angle regime where the induced partition is empirically insensitive to small perturbations. We treat $\theta$ as a fixed operating setting rather than a tuned hyperparameter; alternative choices within the same stable regime act as equivalent representatives and do not materially change the extracted partition statistics under the stated preprocessing and tessellation settings.

Taken together, these limitations delineate the intended use of \textsc{STEP-Parts}: a deterministic, geometry-native reference for CAD instance partitioning that preserves B-Rep analytic structure, while acknowledging that projection to discretized carriers can introduce rare outliers under extreme meshing conditions.

\section{Discussion and Conclusions}
We introduced \textsc{STEP-Parts}, a deterministic CAD-to-supervision toolchain that derives topology-aware geometric instance labels directly from raw STEP B-Reps. The construction operates on intrinsic B-Rep entities---faces, edges, and analytic surface classes---and groups adjacent faces only when they are analytically compatible and locally tangent-continuous. The resulting face-level partition is transferred to a tessellated carrier by retaining, for each triangle, the index of its source B-Rep face, yielding projected STEP-Part labels together with primitive labels and auxiliary metadata. In this way, \textsc{STEP-Parts} provides a reproducible instance-level decomposition that avoids semantic taxonomies and manual annotation while remaining directly usable in downstream learning and evaluation pipelines. Empirical validation on ABC shows that the same-primitive dihedral structure admits a stable low-angle operating regime, that the induced partition remains stable under reasonable changes in tessellation, and that the resulting labels serve as a stronger geometric reference than a mesh-based alternative under the evaluated settings. Two independent downstream probes further indicate that \textsc{STEP-Parts} provides useful supervision for both a shape-level implicit reconstruction--segmentation model and a dataset-level point-based backbone, including a controlled PTv3 probe in which STEP-based supervision improves held-out test performance over PartField-derived labels.

\bibliographystyle{plainnat}
\bibliography{mybibfile}

\end{document}